\documentclass[reprint, showpacs, amsmath, amssymb, aps]{revtex4-1}
\usepackage{mathrsfs}
\usepackage{amsfonts}
\usepackage{amsmath,amsfonts,amssymb,mathrsfs,bm}
\usepackage{verbatim,psfrag,times,CJK}
\usepackage{graphicx,graphics,color,epsfig}
\usepackage{float}
\usepackage{flafter}
\usepackage{subeqnarray}
\usepackage{cases}
\usepackage{amsmath, upgreek, amssymb, graphicx, subfigure, paralist, dsfont}
\usepackage[colorlinks, linkcolor=blue, citecolor=blue, urlcolor=blue, breaklinks=true]{hyperref}

\definecolor{RED}{rgb}{1,0,0}
\newcommand{\ket}[1]{\left\vert{#1}\right\rangle}

\begin{document}

\title{Collective dynamics and entanglement of two atoms embedded into negative index material}

\author{Wei Fang}
\author{Gaoxiang Li}
\email{gaox@phy.ccnu.edu.cn}
\affiliation{Department of Physics, Huazhong Normal University, Wuhan 430079, P. R. China}
\author{Zbigniew Ficek}
\affiliation{The National Center for Applied Physics, KACST, P.O. Box 6086, Riyadh 11442, Saudi Arabia}

\begin{abstract}
We study the dynamics of two two-level atoms embedded near to the interface of paired meta-material slabs, one of negative permeability and the other of negative permittivity. The interface behaves as a plasmonic waveguide composed of surface-plasmon polariton modes. It is found that significantly different dynamics occur for the resonant and an off-resonant couplings of the plasma field to the atoms. In the case of the resonant coupling, the plasma field does not appear as a dissipative reservoir to the atoms. We adopt the image method and show that the dynamics of the two atoms are completely equivalent to those of a four-atom system. Moreover, two threshold coupling strengths exist, one corresponding to the strength of coupling of the plasma field to the symmetric and the other the antisymmetric mode of the two-atom system. The thresholds distinguish between the non-Markovian (memory preserving) and Markovian (memoryless) regimes of the evolutions that different time scales of the evolution of the memory effects and entanglement can be observed. The Markovian regime is characterized by exponentially decaying whereas the non-Markovian regime by sinusoidally oscillating contributions to the evolution of the probability amplitudes. The solutions predict a large and long living entanglement mediated by the plasma field in both Markovian and non-Markovian regimes of the evolution. We also show that a simultaneous Markovian and non-Markovian regime of the evolution may occur in which the memory effects exist over a finite evolution time.
In the case of an off-resonant coupling of the atoms to the plasma field, the atoms interact with each other by exchanging virtual photons which results in the dynamics corresponding to those of two atoms coupled to a common reservoir. In addition, the entanglement is significantly enhanced under the off-resonant coupling.
\end{abstract}


\maketitle

\section{Introduction}

The radiative properties of emitters (e.g. atoms or quantum dots) located inside a dielectric or conducting material can be significantly modified compared to those in vacuum. The modification results from the variation of the density of modes of the EM field which can be adjusted by changing the geometric shape or space period structure of a material~\cite{S.M.Barnett,sk99,W.Lukosz}. The radiative properties of emitters can also be modified by locating  the atoms close to the surface of a dielectric or conducting material~\cite{AG.Tudela,D.E.Chang,V.Karanikolas,J.M.Wylie,lg85,S.M.Dutra,cs09,rb13,sb15}. In this case, the modification results from the presence of surface EM modes known as plasmon guided (PG) field~\cite{Y.Luo,D.E.Beck,W.L.Barnes,R.Ruppin,E.R-G,I.Dolev,P.A.Huidobro}. A new category of materials has been proposed, so-called meta-materials~\cite{H.K.Yuan,S.Linden,D.R.Smith,zz06}, characterized by specifically designed geometrical structures which drastically modify the density of the EM modes, so the field propagation, also yielding to the PG field~\cite{M.Beruete,B.Stein,Y.Li,G.Dolling,M.Cuevas}. Owing to the high local density of modes, emitters may interact strongly with the surface field which affects their radiative properties~\cite{G.X.Li,H.T.Dung,X.D.Zeng,J.Barthes,A.V.Akimov,K.J.Russell,Q.Cheng}. For example, when quantum dots are placed at a distance about several tens of nanometers above two dimensional metal surface, strong coupling could be generated between the quantum dots and the collective mode reflected in the presence of the Rabi oscillations~\cite{A.G.Tudela}. In the structure composed of zero index and left hand materials, maximum quantum interference and a suppression of the atomic decay rate can be achieved between Zeeman levels due to the anisotropy of the EM modes~\cite{G.Song}. It has been shown that by changing the strength of the driving field and adjusting position of the quantum dot, the plasma mode on the surface of a metal-nanoparticles material can introduce asymmetrical features into the spectrum~\cite{R.C.Ge}.

The study of entanglement between distant atoms and controlled the transmission of information between them are vital to the development of quantum information technology~\cite{C.Bennett,sharoche}. A key model to investigate the creation and storage processes of entanglement often follows with a system composed of two-level quantum emitters~\cite{J.Majer,A.Laucht}. When atoms coupled to the same PG field, the incoherent spontaneous exchange of photons could occur and results in collective damping~\cite{D.Dzsotjan,H.X.Zheng}. It has been revealed when the decay through one of the collective states is deeply depressed, long lived entanglement of the system could be achieved, which only depends on the distance between atoms~\cite{gm11}. By applying nanowire structure, the PG field could be well guided and thus entanglement between quantum dots still exists at several vacuum wave lengths~\cite{G.Y.Chen}. Xu {\it et al.}~\cite{J.P.Xu} have shown that the entanglement between two atoms can exist over distances much larger than the resonant wavelength if the space between the atoms is filled with a thin membrane made of single negative ($\varepsilon<0$ or $\mu<0$) and left hand materials. However, most of the studies are focus on metal-dielectric structure, and the entanglement only maintains for a small time scale.

\begin{figure}[t]
\centering\includegraphics[width=8.5cm,keepaspectratio,clip]{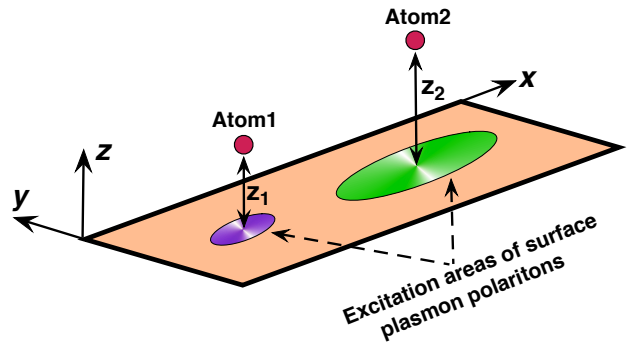}
\caption{(Color online) Two atoms located at distances $z_{1}$ and $z_{2}\, (z_{2}>z_{1})$ close to the interface between two meta-materials. Each atom excites the SPP propagating along the interface. The excitation size of the SPP depends on the distance of the atom from the interface and polarization of its dipole moment. As illustrated by green and purple ellipses, the excitation area of the surface plasmon polaritons decreases with an decreasing distance $z_{i}$ and their shapes extend more in the $x$ direction than in the $y$ direction if the atomic dipole moments are polarized in the $x-z$ plane.}
\label{fig1a}
\end{figure}

Corresponding similarities should also then be expected in the dynamics of atoms located close to the interface between two meta-materials. In this paper, we present an analytical treatment of the dynamics of two independent atoms located close to the interface of two meta-materials, one of a negative permeability (MN) and the other of a negative permittivity (EN). We assume that the atoms are located in the MN material. Surface plasmon polaritons (SPP), nonradiative electromagnetic excitations associated with charge density waves propagating along the interface are generated. The SPP propagate in the $x$- and $y$-directions along the interface between the meta-materials, and rapidly decay in the $z$-direction.  As illustrated in figure~\ref{fig1a}, two atoms located close to the interface can excite the SPP and the excitation depends on the distance of the atoms from the interface and the polarization of the atomic dipole moments. For a polarization of both atomic dipole moments in the $x-z$ plane, the atoms could excite the SPP which would effectively propagate in the $x$ direction. In other words, the polarization of the atomic dipole moments determines the direction of propagation of the SPP on the interface. Hence, the interface would behave as a directional guiding plasma field mode propagating in the $x$-direction, formally analogy of a plasmonic waveguide~\cite{gm11,A.Alu,hg16}. We shall demonstrate below that then it would be possible to achieve a strong interaction between two atoms located close to the interface through their coupling to the SPP.

The mathematical approach we adopt here is based on the Green's function method~\cite{M.S.Tomas}. Our focus is on how the plasma field induced at the interface between the materials changes the dynamics of the atoms, in particular, the population transfer and entanglement. The remarkably simple analytical expressions are derived for the probability amplitudes valid for an arbitrary initial state, arbitrary strengths of the coupling constants of the atoms to the plasma field, and arbitrary distances between the atoms. We find a number of interesting general results. In the first place, we distinguish two different time scales of the evolution of the atomic states, one corresponding to the evolution of the collective symmetric state and the other to the antisymmetric state.
Secondly, we find a threshold behavior of the coupling constants which separate the non-Markovian behavior of the system from the Markovian one~\cite{H.P.Breuer}. The Markovian evolution is usually attributed to a weak coupling of an atom to the field. We show that the collective effects may result in the Markovian evolution even in the limit of a strong coupling of the atoms to the field. Inversely, a non-Markovian evolution can be seen even in the regime of a weak coupling. Thirdly, we find that the plasma field does not appear as a common reservoir to the atoms. In order to explain the behavior of the atoms we adopt the image method and show that the dynamics of the two atoms are completely equivalent to those of a four-atom system.
Finally, we consider the case in which the plasma field frequency is off-resonant with the atoms and find that in this case the dynamics resemble those of two atoms coupled to a common reservoir. The atoms interact through the exchange of virtual photons resulting in the absence of the images.

The plan of this paper is as follows. In section~\ref{sec2} we introduce the model and present the explicit analytic expressions for the time dependence of the probability amplitudes of the atoms. Detailed dynamics of the atoms are studied in section~\ref{sec3}. We assume that a single excitation is present initially in the system and demonstrate how the evolution of the system can be simply understood in terms of the evolution of the atoms and their corresponding images. We then demonstrate in section~\ref{sec4} the collective behavior of the atoms in both Markovian and non-Markovian regimes of the evolution. Entangled properties of the atoms are discussed in section~\ref{sec5}, where we calculate the concurrence for different initial states and different coupling strengths of the atoms to the plasma field. The effect of an off-resonant coupling of the atoms to the plasma field on the collective dynamics and entanglement is discussed in section~\ref{sec6}. We summarize our results in section~\ref{sec7}. The paper concludes with two Appendices in which we give details of the derivation of the integro-differential equations for the probability amplitudes and the calculations of the integral kernels. Both longitudinal and transverse parts of the Green function are considered in the evaluation of the kernels.

\section{Atoms interacting with the plasma field}\label{sec2}

Perfectly conducting materials are known to generate a strong PG field which is refined in a short regime near the surface~\cite{J.B.Pendry,A.Huck,RRuppin}. A plasma field can also be generated at the surface of a meta-material with either negative permittivity ($\varepsilon<0$) or negative permeability ($\mu<0$). However, the density of the plasma field near surface only originates from one or several discrete modes, which can be derived by applying the continuous conditions on the boundaries. Recently, Tan {\it et al.}~\cite{W.Tan} have shown that the density of the EM modes can be significantly enhanced at the interface of two meta-materials, one with negative $\varepsilon$ and the other with negative $\mu$. Especially, when the materials are perfectly paired, i.e. $\varepsilon_1 = -\varepsilon_2$ and $\mu_1 = -\mu_2$, the effective permittivity and permeability, defined as
\begin{eqnarray}
\varepsilon_r = \frac{d_1\varepsilon_{1} + d_2\varepsilon_2}{d_1+d_2} ,\quad
\mu_r = \frac{d_1\mu_{1} + d_2\mu_2}{d_1+d_2} ,
\end{eqnarray}
are both zero when the slabs have the same thickness, $d_1=d_2$. In this case, the band gap disappears and the density of modes of the EM field becomes continuous. This means that a large density of the modes exists at the interface between the two perfectly paired negative meta-material slabs, and can be treated as optical topological material~\cite{X.Huang}.

\begin{figure}[h]
\centering\includegraphics[width=8.5cm,keepaspectratio,clip]{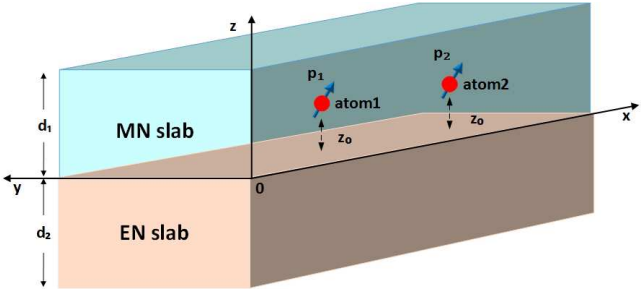}
\caption{(Color online) Geometry of the system. The $z$ axis is taken normal to the interface between MN and EN slabs with its origin at the interface. The slabs have thickness $d_1$ and $d_2$, respectively, and are assumed to have infinite extents in the $xy$ plane. Two atoms are embedded in the MN slab at fixed positions $(x_{1},0,z_{0})$ and $(x_{2},0,z_{0})$, where $z_{0}$ is the distance of the atoms from the interface between the materials. The atomic transition dipole moments ${\bf p}_{1}$ and ${\bf p}_{2}$ are parallel to each other and oriented in the $x-z$ plane.}
\label{fig1}
\end{figure}

We consider a system composed of two identical atoms located at a distance $z_{0}$ from the interface between two different negative index material slabs, $\mu$-negative (MN) and $\varepsilon$-negative (EN) slabs, as shown in figure~\ref{fig1}. We assume that the atoms are located in the MN slab and the distance between the atoms, $x_{21}=x_{2}-x_{1}$, is large compared to the atomic wavelength, $x_{21}\gg \lambda_{a}$, so there is no direct interaction between the atoms. Each atom is represented by its ground state $\ket{g_{j}}$, an excited state $\ket{e_{j}}\, (j=1,2)$, the atomic transition frequency $\omega_{a}$, and the atomic transition dipole moment $\vec{p}_{j}$.

The atoms interact with an electromagnetic field via a dipole interaction according to the Hamiltonian~\cite{S.Y.Buhmann}
\begin{eqnarray}
\hat{H} = \hat{H}_{0} + \hat{H}_{I} ,\label{h1}
\end{eqnarray}
where
\begin{eqnarray}
\hat{H}_{0} &=& \frac{1}{2}\hbar\omega_{a}(\hat{\sigma}_{z1} + \hat{\sigma}_{z2}) \nonumber\\
&+&\hbar \sum_{\lambda=e,m}\int{d{\bf r}}\int_{0}^{\infty}d\omega\, \omega\, \hat{\bf f}_\lambda^{\dag}({\bf r},\omega)\hat{{\bf f}}_\lambda({\bf r},\omega) \label{h2}
\end{eqnarray}
is the unperturbed Hamiltonian of the atoms and the field, and
\begin{eqnarray}
\hat{H}_{I} = - \sum_{j=1,2}\left[{\bf p}_{j}\cdot\int_0^\infty d\omega\hat{{\bf E}}^{(+)}({\bf r}_j,\omega)\hat{\sigma}^{\dag}_{j} +\textrm{H.c.}\right] \label{h3}
\end{eqnarray}
is the interaction of the atoms with the field. Here, $\hat{\bf f}_\lambda^{\dag}({\bf r},\omega)$ and $\hat{{\bf f}}_\lambda({\bf r},\omega)$ are the creation and annihilation operators which can be viewed as collective excitations of the electromagnetic field, $\lambda=e$ represents noise polarization of the EN material, $\lambda=m$ represents noise magnetization of the MN material~\cite{sb15,B.Huttner}, $\hat{\sigma}^{\dagger}_{j} (\hat{\sigma}_{j})$ and $\hat{\sigma}_{z}$ are the raising (lowering) and the energy difference operators of atom $j$. The positive frequency part of the electric field operator at the position ${\bf r}_{j}$ of the $j$th atom is given by
\begin{eqnarray}
\hat{{\bf E}}^{(+)}({\bf r}_{j},\omega) &=& i\sqrt{\frac{\hbar}{\pi\varepsilon_{0}}}\frac{\omega}{c} \nonumber\\
&\times& \int d{\bf r}  \left\{\frac{\omega}{c}\sqrt{\Im[\varepsilon({\bf r},\omega)]} \stackrel{\leftrightarrow}{\bf G}({\bf r}_{j},{\bf r},\omega)\cdot \hat{{\bf f}}_e({\bf r},\omega)\right. \nonumber\\
&+& \left. \!\sqrt{-\Im[\kappa({\bf r},\omega)]}{\bf \nabla}\times\!\stackrel{\leftrightarrow}{\bf G}\!\!({\bf r}_{j},{\bf r},\omega)\!\cdot\!\hat{{\bf f}}_m({\bf r},\omega)\!\right\},
\end{eqnarray}
where $\Im[\varepsilon({\bf r},\omega)]$ is the imaginary part of permittivity, $\Im[\kappa({\bf r},\omega)]$ is the imaginary part of reciprocal of permeability ($\kappa({\bf r},\omega)=1/\mu({\bf r},\omega)$), respectively, and $\stackrel{\leftrightarrow}{\bf G}({\bf r}_{j},{\bf r},\omega)$ is the Green tensor of the field, which  characterizes the density of the field modes at the location ${\bf r}_{j}$ of atom.

For the permittivity and permeability of the slabs, we assume that $\varepsilon_{1}$ and $\mu_{2}$ are positive constants, but $\varepsilon_{2}$ and $\mu_{1}$ are negative and strongly depend on frequency of electromagnetic field which have following forms~\cite{H.T.Jiang,R.A.Shelby}
\begin{eqnarray}
\frac{\varepsilon_{2}(\omega)}{\varepsilon_0} &=& 1+\frac{\omega_{ep}^2}{{\omega_{eo}}^2-{\omega}^2-i\omega\gamma_e} ,\nonumber\\
\frac{\mu_{1}(\omega)}{\mu_0} &=& 1+\frac{\omega_{mp}^2}{{\omega_{mo}}^2-{\omega}^2-i\omega\gamma_m} ,\label{e1}
\end{eqnarray}
where $\omega_{ep}$ and $\omega_{mp}$ are plasma frequencies of the electric and magnetic materials, respectively, $\omega_{eo}$ and $\omega_{mo}$ are resonance frequencies of the materials, and $\gamma_{e}, \gamma_{m}$ are dissipation (losses) parameters of the materials. For clarity of the notation we have omitted the spatial argument. It is clear from Eq.~(\ref{e1}) that in the frequency region above the resonance, $\omega >\omega_{eo}\, (\omega >\omega_{mo})$ and $\omega<\omega_{ep}\, (\omega<\omega_{mp})$, the EN (MN) slab is a single-negative material. Thus, the structure of a meta-material can be designed~\cite{I.Bergmair,N.T.Tung}.

To study the dynamics of the atoms, we consider the wave function of the system whose the time evolution is governed by the Schr\"{o}dinger equation
\begin{eqnarray}
i\hbar\frac{\partial}{\partial t}\ket{\Psi(t)} = \hat{H}_{I}\ket{\Psi(t)} .\label{es}
\end{eqnarray}
If the field was initially at $t=0$ in the vacuum state and the atoms shared a single excitation, the wave function of the system at time $t>0$, written in the interaction picture is of the form
\begin{align}
\ket{\Psi(t)} &= C_1(t)e^{-i\omega_{a}t}\ket{\{0\}}\ket{e_1,g_2} \nonumber\\
&+ C_2(t)e^{-i\omega_{a}t}\ket{\{0\}}\ket{g_1,e_2} + \sum_{\lambda=e,m} \int_{0}^{\infty} d\omega\, e^{-i\omega t} \nonumber\\
&\times \int d{\bf r}\, {\bf C}_\lambda({\bf r},\omega,t)\cdot\ket{{\bf 1_\lambda}({\bf r},\omega)}\ket{g_1,g_2} ,
\end{align}
where $C_{1}(t)$ is the probability amplitude of the state in which atom $1$ is in its excited state $\ket{e_{1}}$, atom $2$ is in the ground state $\ket{g_{2}}$, and the field is in the vacuum state $\ket{\{0\}}$, $C_{2}(t)$ is the probability amplitude of the state in which atom $1$ is in its ground state $\ket{g_{1}}$, atom $2$ is in the excited state $\ket{e_{2}}$, and the field is in the vacuum state $\ket{\{0\}}$, and ${\bf C}_{\lambda}({\bf r},\omega,t)$ is the probability amplitude of the state in which both atoms are in their ground states, $\ket{g_{1}}, \ket{g_{2}}$, and there is an excitation of the medium-assisted field $\ket{{\bf 1_\lambda}({\bf r},\omega)}\equiv \hat{\bf f}_\lambda^\dag({\bf r},\omega)\ket{\{0\}}$.

With the interaction (\ref{h3}), the Schr\"{o}dinger equation (\ref{es}) transforms into four coupled equations of motion for the probability amplitudes. When the amplitudes $C_\lambda({\bf r},\omega,t)$ are eliminated we arrive, as shown in Appendix A, into two coupled integro-differential equations for the probability amplitudes of the atoms
\begin{align}
\dot{C}_1(t) &= \int^t_{0}\!dt^{\prime} K_{11}(t,t^{\prime}) C_1(t')\!+\!\int^t_{0}\!dt^{\prime} K_{12}(t,t^{\prime}) C_2(t') ,\label{e11u}\\
\dot{C}_2(t) &= \int^t_{0} dt^{\prime} K_{22}(t,t^{\prime}) C_2(t')\!+\!\int^t_{0} dt^{\prime} K_{21}(t,t^{\prime}) C_1(t') ,\label{e12u}
\end{align}
in which $K_{ii}(t,t^{\prime})$ is the integral kernel determined by the imaginary part of the one-point Green tensor, $\stackrel{\leftrightarrow}{\bf G}\!({\bf r}_{i},{\bf r}_{i},\omega)$, whereas $K_{ij}(t,t^{\prime}), \, i\neq j$ is the integral kernel determined by the two-point Green tensor, $\stackrel{\leftrightarrow}{\bf G}\!({\bf r}_{i},{\bf r}_{j},\omega)$.

The kernels can be evaluated explicitly, and straightforward but lengthly calculations (for details see Appendix B) lead to the following explicit expressions
\begin{align}
K_{11}(t,t^{\prime}) &= K_{22}(t,t^{\prime}) = -\Omega^{2}_{0}e^{-\left(\frac{1}{2}\gamma +i\delta\right)(t-t')} ,\label{ek1}\\
K_{12}(t,t^{\prime}) &= K_{21}(t,t^{\prime}) = -\Omega^{2}_{0}U(x_{21},z_{0}) e^{-\left(\frac{1}{2}\gamma +i\delta\right)(t-t')} ,\label{ek2}
\end{align}
where $\delta =\omega_{s}-\omega_{a}$ is the detuning of the atomic transition frequency from the plasma field frequency,
\begin{eqnarray}
\Omega_{0} = \left\{\frac{\omega_s \Gamma_{A}\!\left[3+4\pi^{2}\!\left|\Re\,[\mu_{1}(\omega_s)]\right|\!\left(2z_{0}/\lambda_{s}\right)^2\right]}{64\pi^{3}(2z_{0}/\lambda_{s})^3}\right\}^{1/2} \label{om0}
\end{eqnarray}
is the coupling strength of the atoms to the surface plasma field, and
\begin{align}
U(x_{21},z_{0}) &= F\!\left[\frac{1}{2},1,2;-\frac{x_{21}^2}{(2z_{0})^2}\right] \nonumber\\
&+ \frac{1}{3+4\pi^{2}\!\left|\Re\,[\mu_{1}(\omega_s)]\right|\!\left(2z_{0}/\lambda_{s}\right)^{2}} \nonumber\\
&\times \left\{F\!\left[\frac{3}{2},2,2;-\frac{x_{21}^2}{(2z_{0})^2}\right]\!+\!2F\!\left[\frac{3}{2},2,1;-\frac{x_{21}^2}{(2z_{0})^2}\right]\right. \nonumber\\
&-\left. 3F\!\left[\frac{1}{2},1,2;-\frac{x_{21}^2}{(2z_{0})^2}\right]\right. \nonumber\\
&-\left. 3\frac{x_{21}^2}{(2z_{0})^2}F\left[\frac{5}{2},3,3;-\frac{x_{21}^2}{(2z_{0})^2}\right]\right\} \label{u}
\end{align}
determines the strength of the interaction between the atoms resulting from the coupling of the atoms with the same plasma field.
Here, $\Gamma_{A} = \omega_{a}^{3}p_{a}^{2}/(3\varepsilon_{0}\pi\hbar c^{3})$ is the spontaneous emission rate of the atoms in free space, assumed the atoms are identical, $p_{a}\equiv p_{1}=p_{2}$.
\begin{figure}[h]
\centering\includegraphics[width=9cm,keepaspectratio,clip]{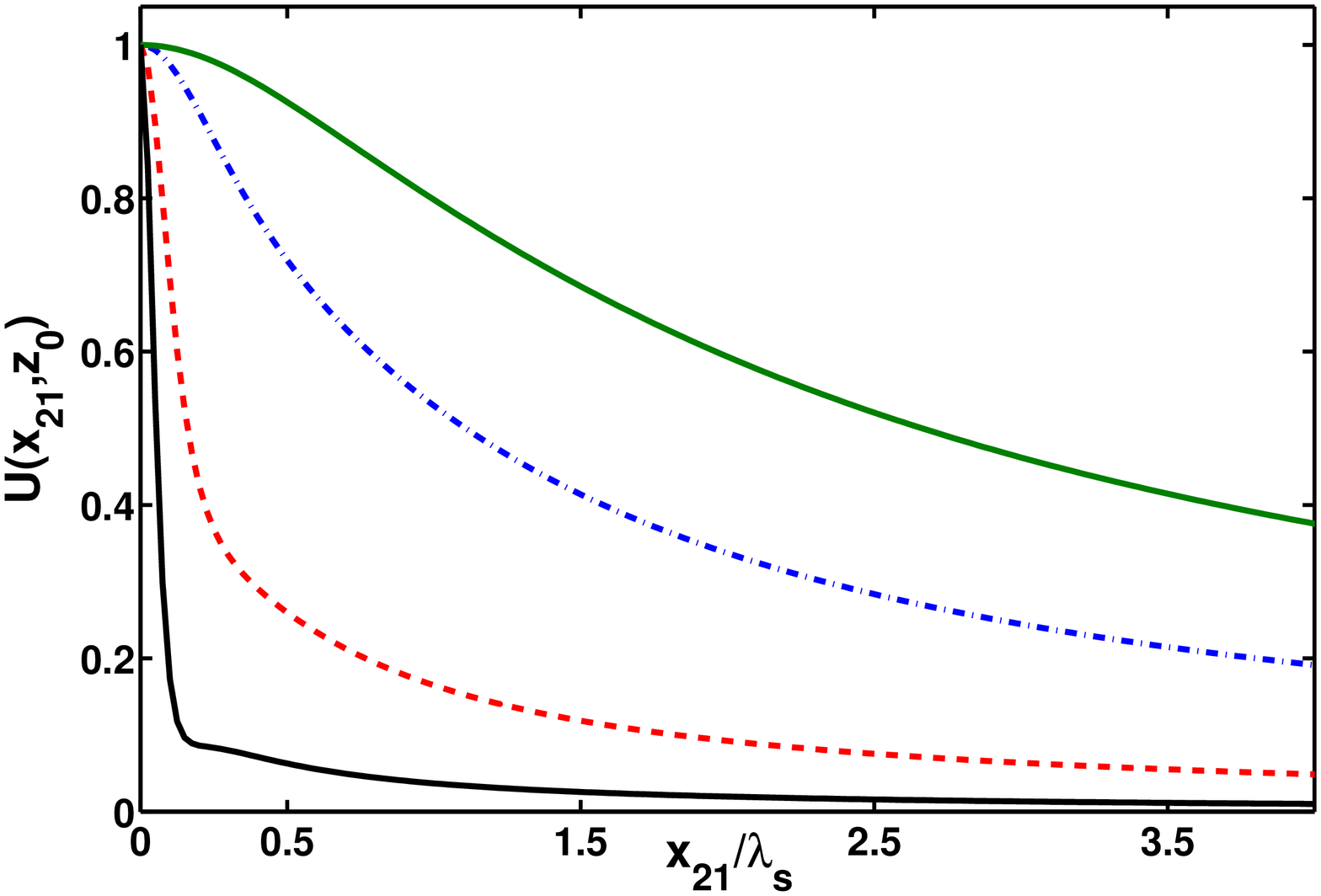}
\caption{(Color online) Dependence of $U(x_{21},z_{0})$ on separation between the atoms $x_{21}/\lambda_{s}$ for several different distances of the atoms from the interface: $z_{0}=0.05\lambda_{s}$ (solid black line), $z_{0}=0.1\lambda_{s}$ (dashed red line), $z_{0}=0.25\lambda_{s}$ (dashed-dotted blue line), and $z_{0}=0.5\lambda_{s}$ (solid green line).}
\label{fig4}
\end{figure}

The function $U(x_{21},z_{0})$ depends on the separation between the atoms, $x_{21}$, and also their distance $z_{0}$ from the interface. It determines the strength of the coupling between the atoms. In the limit of $x_{21}\gg\lambda_{s}$, $U(x_{21},z_{0})\approx 0$, while for $x_{21}\ll\lambda_{s}$, $U(x_{21},z_{0})\approx 1$. Thus, for large $x_{21}$, the effects of the coupling between the atoms become negligible and the atoms evolve independently.  Notice that at $x_{21}=0$ the function $U(x_{21},z_{0})$ is always unity independent of the value of $z_{0}$, the distance of the atoms from the interface. However, the variation of $U(x_{21},z_{0})$ with $x_{21}$ depends strongly on $z_{0}$. This is illustrated in figure~\ref{fig4} which shows $U(x_{21},z_{0})$  as a function of  $x_{21}$ for several different values of $z_{0}$. It is seen that for large $z_{0}$, the function $U(x_{21},z_{0})$ varies slowly with $x_{21}$. In that case, the indirect coupling between the atoms which is provided by the plasma field is effectively quite strong even at large separations. On the other hand, for small $z_{0}$, $(z_{0}\ll \lambda_{s})$, the function $U(x_{21},z_{0})$ is different from zero only over very small distances $x_{21}$ and decays rapidly to zero as $x_{21}$ increases. 

Thus, a strong coupling of the atoms to the plasma field {\it destroys} the collective behavior of the atoms. In the physical terms, the location of the atoms at a small distance $z_{0}$ from the interface leads to a strong spatial confinement (localization) of the surface plasmon fields around $x_{1}$ and $x_{2}$ resulting in a weak overlap of the surface fields produced by the atoms.

\section{Collective dynamics of the atoms coupled to surface plasma}\label{sec3}

Having available the explicit expressions for the integral kernels, we can now solve Eqs.~(\ref{e11u}) and (\ref{e12u}) and study the time evolution of the atomic system.
If we introduce symmetric and antisymmetric combinations of the probability amplitudes, $C_{s} =(C_{1}+C_{2})/\sqrt{2}$ and $C_{a}=(C_{1}-C_{2})/\sqrt{2}$, corresponding to collective symmetric and antisymmetric states of the two-atom system, we readily find that Eqs.~(\ref{e11u}) and (\ref{e12u}) simplify to
\begin{eqnarray}
\dot{C}_{s}(t) &=& \int_0^tdt^{\prime} K_{s}(t,t^{\prime}) C_{s}(t') ,\label{a3}\\
\dot{C}_{a}(t) &=& \int_0^tdt^{\prime} K_{a}(t,t^{\prime})C_{a}(t') ,\label{a4}
\end{eqnarray}
where
\begin{eqnarray}
K_{s}(t,t^{\prime}) &= -\Omega^{2}_{s}e^{-(\frac{1}{2}\gamma +i\delta)(t-t')} , \nonumber\\
K_{a}(t,t^{\prime}) &= -\Omega^{2}_{a}e^{-(\frac{1}{2}\gamma +i\delta)(t-t')} .\label{a5}
\end{eqnarray}
Here, $\Omega^{2}_{s} = \Omega_{0}^{2}[1+U(x_{21},z_{0})]$ and $\Omega^{2}_{a} = \Omega_{0}^{2}[1-U(x_{21},z_{0})]$ are coupling strengths of the symmetric and antisymmetric states to the plasma field, respectively. Clearly, the coupling strengths of the collective states are altered by the atomic interaction with $\Omega_{s}$ enhanced and $\Omega_{a}$ reduced by $U(x_{21},z_{0})$.
This may have an interesting effect on the dynamics of the atoms that at small distances between the atoms, at which $U(x_{21},z_{0})\approx 1$, the antisymmetric state could be completely decoupled from the interaction with the plasma field leaving only the symmetric state to be strongly coupled to the field.

It is seen from Eqs.~(\ref{a3}) and (\ref{a4}) that the equations of motion for the probability amplitudes $C_{s}(t)$ and $C_{a}(t)$ are independent of each other and are similar in form. We therefore need to obtain the solution for $C_{s}(t)$, and then the solution for $C_{a}(t)$ can be obtained simply by replacing $\Omega_{s}$ by $\Omega_{a}$.
From Eq.~(\ref{a5}) it follows that the kernel $K_{s}(t,t^{\prime})$ is a function only of the time difference $t-t^{\prime}$. Therefore, the integro-differential equation (\ref{a3}) can be solved exactly by Laplace transformation. Thus if
\begin{align}
\int_{0}^{\infty}dt K_{s}(t,t^{\prime})e^{-pt} = K_{s}(p) ,\ \int_{0}^{\infty}dt C_{s}(t)e^{-pt} = C_{s}(p) ,\label{a9}
\end{align}
we get from Eq.~(\ref{a3})
\begin{eqnarray}
pC_{s}(p) - C_{s}(0) = K_{s}(p)C_{s}(p) ,\label{a10}
\end{eqnarray}
or
\begin{eqnarray}
C_{s}(p) = \frac{C_{s}(0)}{p -K_{s}(p)} = \frac{C_{s}(0)(p+\frac{1}{2}\gamma +i\delta)}{p(p+\frac{1}{2}\gamma +i\delta) +\Omega^{2}_{s}} .\label{a11}
\end{eqnarray}

By inverse Laplace transformation we then have the result
\begin{align}
C_{s}(t) &= \frac{1}{2}C_{s}(0)e^{-\frac{1}{2}i\delta t} \left\{\left[1+\frac{(\gamma + 2i\delta)}{4\tilde{\Omega}_{s}}\right]e^{-\left(\frac{1}{4}\gamma -\tilde{\Omega}_{s}\right) t} \right. \nonumber\\
&+\left. \left[1-\frac{(\gamma\!+\!2i\delta)}{4\tilde{\Omega}_{s}}\right]e^{-\left(\frac{1}{4}\gamma +\tilde{\Omega}_{s}\right) t}\right\} ,\label{a42}
\end{align}
where $\tilde{\Omega}_{s} = \sqrt{\frac{1}{16}(ue^{i\theta})^{2} -\Omega_{s}^{2}}$ is the effective Rabi frequency of the interaction of the atoms with the plasma field, $u=\sqrt{\gamma^{2}+4\delta^{2}}$ and $\theta =\arctan(2\delta/\gamma)$. Similarly, for $C_{a}(t)$, we get
\begin{align}
C_{a}(t) &= \frac{1}{2}C_{a}(0)e^{-\frac{1}{2}i\delta t} \left\{\left[1+\frac{(\gamma\!+\!2i\delta)}{4\tilde{\Omega}_{a}}\right]e^{-\left(\frac{1}{4}\gamma -\tilde{\Omega}_{a}\right) t}\right. \nonumber\\
&+\left. \left[1-\frac{(\gamma\!+\!2i\delta)}{4\tilde{\Omega}_{a}}\right]e^{-\left(\frac{1}{4}\gamma +\tilde{\Omega}_{a}\right) t}\right\} ,\label{a43}
\end{align}
where $\tilde{\Omega}_{a} = \sqrt{\frac{1}{16}(ue^{i\theta})^{2} -\Omega_{a}^{2}}$.

Two important features of the results (\ref{a42}) and (\ref{a43}) should be noted. Firstly, the effective Rabi frequencies $\tilde{\Omega}_{s}$ and $\tilde{\Omega}_{a}$ exhibit a threshold effect that depending upon $\Omega_{s(a)}<u/4$ or $\Omega_{s(a)}>u/4$, the Rabi frequencies can be either purely real or purely imaginary. In the other words, the time evolution of the probability amplitudes could be either exponential or sinusoidal. Secondly, we note that below threshold the time evolution of both $C_{s}(t)$ and $C_{a}(t)$ involves two decaying exponentials with a reduced (subradiant) decay constant, $\frac{1}{4}\gamma -\tilde{\Omega}_{s(a)}$, and an enhanced (superradiant) decay constant, $\frac{1}{4}\gamma +\tilde{\Omega}_{s(a)}$. The involvement of the fast and slow decay rates in the evolution of both superradiant and subradiant states seems to contradicts our expectation since, according to Dicke~\cite{rd56} (see also Refs.~\cite{rl70,sl10,zf9}), each of the collective amplitudes of a two atom system should decay with a single rate: The symmetric superposition $C_{s}(t)$ should decay with the fast (superradiant) rate while $C_{a}(t)$ should decay with the slow (subradiant) rate.
\begin{figure}[h]
\centering\includegraphics[width=8cm,keepaspectratio,clip]{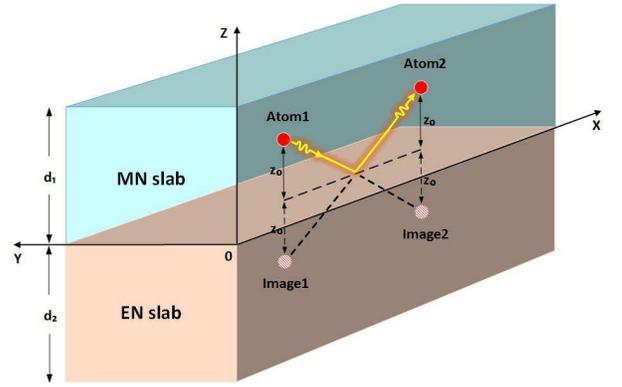}
\caption{(Color online) Two atoms located at a distance $z_{0}$ from the interface between two materials and their images located at a distance $z_{0}$ behind the interface. The atoms are not directly coupled to each other, but can be coupled by the radiation reflected from the interface. A photon emitted by atom $1$ and reflected from the interface towards atom $2$ can be viewed as being emitted from the image of the atom~$1$.}
\label{fig3}
\end{figure}

A qualitative understanding of the involvement of both fast and slow decay rates in the evolution of $C_{s}(t)$ and $C_{a}(t)$ may be obtained by considering the interaction of the atoms with the plasma field as the interaction with images of the atoms located at a distance $z_{0}$ behind the interface. This is illustrated in Fig.~\ref{fig3}, which shows that the two-atom system interacting with the surface plasma field can be seen as a four-qubit system, the two atoms plus two images.
The radiation field emitted by either atom $1$ or atom $2$ and reflected from the interface in the direction normal to the interface can be regarded as the radiation from an image located at a distance $z_{0}$ behind the interface. The radiation field emitted by atom $1$ and reflected from the interface towards atom $2$ can be viewed as being emitted by the image of the atom $1$ located at a distance $z_{0}$ behind the interface.

To study the evolution of the atoms in terms of the interaction with their images, we write Eqs.~(\ref{e11u}) and (\ref{e12u}) in the following form
\begin{align}
\dot{\tilde{C}}_1(t) &= \left(\frac{1}{2}\gamma +i\delta\!\right)\!\tilde{C}_{1}(t) -\Omega^{2}_{0}\!\int^t_{0}\!dt^{\prime}\tilde{C}_1(t') \nonumber\\
&-\Omega^{2}_{0}U(x_{21},\!z_{0})\!\int^t_{0}\!dt^{\prime}\tilde{C}_2(t') ,\label{e11w}\\
\dot{\tilde{C}}_2(t) &= \left(\frac{1}{2}\gamma +i\delta\!\right)\!\tilde{C}_{2}(t) -\Omega^{2}_{0}\!\int^t_{0}\!dt^{\prime}\tilde{C}_2(t')  \nonumber\\
&-\Omega^{2}_{0}U(x_{21},\!z_{0})\!\int^t_{0}\!dt^{\prime} \tilde{C}_1(t') ,\label{e12w}
\end{align}
where $\tilde{C}_1(t)=C_{1}(t)\exp[(\frac{1}{2}\gamma +i\delta)t]$ and $\tilde{C}_{2}(t)=C_{2}(t)\exp[(\frac{1}{2}\gamma +i\delta)t]$.

Following the Fig.~\ref{fig3}, the second term on the right-hand side of Eq.~(\ref{e11w}) may be interpreted as arising from the coupling of the atom $1$ to its image, whereas the third term may be interpreted as arising from the coupling of the atom $1$ to the image of the atom $2$. Thus, we can immediately write Eqs.~(\ref{e11w}) and (\ref{e12w}) as
\begin{align}
\dot{\tilde{C}}_1(t) &= \left(\frac{1}{2}\gamma +i\delta\right)\tilde{C}_{1}(t) +i\Omega_{0}\tilde{C}_{1I}(t) \nonumber\\
&+i\Omega_{0}U(x_{21},z_{0})\tilde{C}_{2I}(t) ,\label{e11z}\\
\dot{\tilde{C}}_2(t) &= \left(\frac{1}{2}\gamma +i\delta\right)\tilde{C}_{2}(t) +i\Omega_{0} \tilde{C}_{2I}(t) \nonumber\\
&+i\Omega_{0}U(x_{21},z_{0})\tilde{C}_{1I}(t) ,\label{e12z}
\end{align}
where
\begin{align}
\tilde{C}_{1I}(t) = i\Omega_{0}\int^t_{0}\!dt^{\prime}\tilde{C}_{1}(t') ,\quad \tilde{C}_{2I}(t) = i\Omega_{0}\int^t_{0} dt^{\prime}\tilde{C}_2(t') \label{e12r}
\end{align}
are the probability amplitudes of the images of the atom $1$ and $2$, respectively.

The equation of motion for the probability amplitudes of the corresponding images are
\begin{eqnarray}
\dot{\tilde{C}}_{1I}(t) &=& i\Omega_{0}\tilde{C}_{1}(t) +i\Omega_{0}U(x_{21},z_{0})\tilde{C}_{2}(t) ,\label{e11s}\\
\dot{\tilde{C}}_{2I}(t) &=& i\Omega_{0} \tilde{C}_{2}(t) +i\Omega_{0}U(x_{21},z_{0})\tilde{C}_{1}(t) .\label{e12s}
\end{eqnarray}
We focus on the evolution of the symmetric combinations of the probability amplitudes, which obey the equations
\begin{eqnarray}
\dot{\tilde{C}}_{s}(t) &=& \left(\frac{1}{2}\gamma +i\delta\right)\tilde{C}_{s}(t) +i\Omega_{s}\tilde{C}_{sI}(t) ,\nonumber\\
\dot{\tilde{C}}_{sI}(t) &=& i\Omega_{s}\tilde{C}_{s}(t) .\label{e29}
\end{eqnarray}
It is then straightforward to show that the solution of Eq.~(\ref{e29}) for $C_{s}(t)$ is of the same form as Eq.~(\ref{a42}). The involvement of the images allows us to write the general solution for $C_{s}(t)$ as a sum of two amplitudes
\begin{align}
C_{s}(t) = \frac{(\gamma\!+\!2i\delta)}{8\tilde{\Omega}_{s}}\!\left[\tilde{D}_{a}(t)\cos\phi +i\tilde{D}_{s}(t)\sin\phi\right]\!e^{-\left(\frac{1}{2}\gamma +i\delta\right)t} ,
\end{align}
where
\begin{align}
\tilde{D}_{s}(t) &= i\tilde{C}_{s}(t)\sin\phi +\tilde{C}_{sI}(t)\cos\phi \nonumber\\
&= iC_{s}(0)e^{\left[\frac{1}{4}(\gamma +2i\delta)-\tilde{\Omega}_{s}\right]t}\sin\phi ,\nonumber\\
\tilde{D}_{a}(t) &= \tilde{C}_{s}(t)\cos\phi -i\tilde{C}_{sI}(t)\sin\phi \nonumber\\
&= C_{s}(0)e^{\left[\frac{1}{4}(\gamma +2i\delta)+\tilde{\Omega}_{s}\right]t}\cos\phi ,
\end{align}
are symmetric and antisymmetric superpositions of the probability amplitudes of the atomic and image states, with
\begin{eqnarray}
\cos^{2}\phi = \frac{1}{2} +\frac{2\tilde{\Omega}_{s}}{(\gamma + 2i\delta)} .\label{c1}
\end{eqnarray}

A similar treatment can be applied to $C_{a}(t)$, which can be written in the form
\begin{align}
C_{a}(t) = \frac{(\gamma + 2i\delta)}{8\tilde{\Omega}_{s}}\!\left[\tilde{G}_{a}(t)\cos\psi +i\tilde{G}_{s}(t)\sin\psi\right]\!e^{-\left(\frac{1}{2}\gamma +i\delta\right)t} ,
\end{align}
where
\begin{align}
\tilde{G}_{s}(t) &= i\tilde{C}_{a}(t)\sin\psi +\tilde{C}_{aI}(t)\cos\psi \nonumber\\
&= iC_{a}(0)e^{\left[\frac{1}{4}(\gamma +2i\delta)-\tilde{\Omega}_{a}\right]t}\sin\psi ,\nonumber\\
\tilde{G}_{a}(t) &= \tilde{C}_{a}(t)\cos\psi -i\tilde{C}_{aI}(t)\sin\psi \nonumber\\
&= C_{a}(0)e^{\left[\frac{1}{4}(\gamma +2i\delta)+\tilde{\Omega}_{a}\right]t}\cos\psi ,
\end{align}
with
\begin{eqnarray}
\cos^{2}\psi = \frac{1}{2} +\frac{2\tilde{\Omega}_{a}}{(\gamma + 2i\delta)} .\label{c2}
\end{eqnarray}
The reason for the presence of both the superradiant and subradiant terms in Eqs.~(\ref{a42}) and (\ref{a43}) is now clear: The superradiant terms are associated with the decay of the symmetric superpositions involving the atomic and image states, $\tilde{D}_{s}(t)$ and $\tilde{G}_{s}(t)$, whereas the subradiant terms are associated with the decay of the antisymmetric superpositions $\tilde{D}_{a}(t)$ and $\tilde{G}_{a}(t)$. The slowest decay rate in the system, $\gamma_{a}^{-}=\frac{1}{4}\gamma -\tilde{\Omega}_{a}$, is the decay rate of the antisymmetric superposition $\tilde{G}_{a}(t)$ whereas the fastest decay rate, $\gamma_{s}^{+}=\frac{1}{4}\gamma +\tilde{\Omega}_{s}$, is the decay rate of the symmetric superposition $\tilde{D}_{s}(t)$.

We may conclude that the interaction of the atoms with the surface plasma field can be viewed as the interaction between the atoms and their corresponding images.

\section{Markovian and non-Markovian regimes of the evolutions}\label{sec4}

We have already seen that different locations of the atoms lead to two different Rabi frequencies $\tilde{\Omega}_{s}$ and $\tilde{\Omega}_{a}$ determining the evolution of the system and though two different time scales of the evolution. The forms of $\tilde{\Omega}_{s}$ and $\tilde{\Omega}_{a}$ show a threshold effect for the Rabi frequencies that depending upon $\Omega_{s(a)}<u/4$ or $\Omega_{s(a)}>u/4$, the time evolution of the probability amplitudes could be either exponential or sinusoidal.
Since $\Omega_{s}\neq \Omega_{a}$, the threshold conditions for the evolution of the symmetric and antisymmetric states do not coincide with each other, and thus we can distinguish between three regions of $\Omega_{s}$ and $\Omega_{a}$: (a) $\Omega_{s}<u/4$ and $\Omega_{a}<u/4$, (b) $\Omega_{s}>u/4$ and $\Omega_{a}<u/4$, (c) $\Omega_{s}>u/4$ and $\Omega_{a}>u/4$. Physically, the threshold values of $\tilde{\Omega}_{s}$ and $\tilde{\Omega}_{a}$ separate what we can identify as the non-Markovian (memory preserved) regime from the Markovian (memoryless) regime of the evolution~\cite{bp02,al14,mx15,ma15}. The case (a), in which $\tilde{\Omega}_{s}$ and $\tilde{\Omega}_{a}$ are real so the Rabi frequencies contribute to the decay rates, that the exponentially decaying amplitudes of the symmetric and antisymmetric states are a manifestation of a Markovian evolution. In the case (b), the dynamics of the system are partly Markovian and partly non-Markovian. The symmetric state undergoes a non-Markovian whereas the antisymmetric state undergoes a Markovian evolutions. In the case (c), the dynamics of the system are fully non-Markovian that the amplitudes of both  symmetric and antisymmetric states undergo an oscillatory evolution, which is a manifestation of a non-Markovian evolution. A non-Markovian evolution is a reversible process characterized by a flow (oscillation) of the information between the atoms and the field, but a Markovian evolution is a irreversible process of a flow (decay) of the information to the field. A possibility to control the non-Markovian dynamics is essential in quantum information technology since it plays a crucial role in preserving quantum memory.

\subsection{Both $\Omega_{s}$ and $\Omega_{a}$ below threshold}

Let us specialize Eqs.~(\ref{a42}) and (\ref{a43}) to the case of exact resonance, $\delta =0$, and first examine the situation when the coupling of the atoms to the plasma field is weak, $\Omega_{0}\ll \gamma$. In this case, both $\tilde{\Omega}_{s}$ and $\tilde{\Omega}_{a}$ could be below threshold. If $\Omega_{s(a)}<u/4$, then $\tilde{\Omega}_{s}$ and $\tilde{\Omega}_{a}$ are real and we see from Eqs.~(\ref{a42}) and (\ref{a43}) that the Rabi frequencies contribute to the damping rates of the probability amplitudes. Physically, this is the kind of behavior corresponding to a Markovian evolution. Since $\Omega_{s}\neq \Omega_{a}$, we see that the weak coupling of the atoms to the plasma field may result in the decay of the probability amplitudes $C_{1}(t)$ and $C_{2}(t)$ with four rates, two enhanced and two reduced rates.
\begin{figure}[h]
\centering\includegraphics[width=9cm,keepaspectratio,clip]{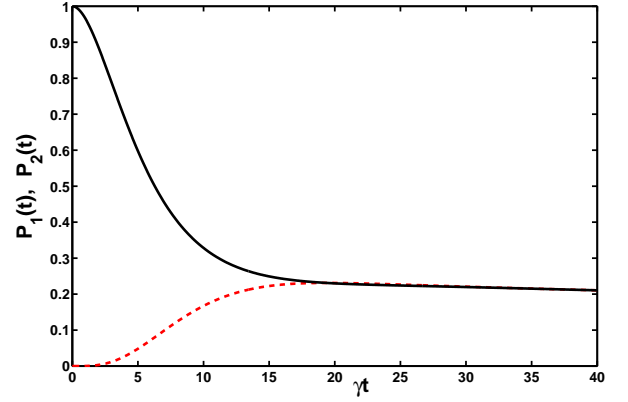}
\caption{(Color online) Time evolution of the populations  $P_{1}(t)=|C_{1}(t)|^{2}$ (black solid line) and $P_{2}(t)=|C_{2}(t)|^{2}$ (red dashed line) for $\delta=0$, $\Omega_{0} = 0.15\gamma$, and $U(x_{21},z_{0})=0.95$, corresponding to both $\Omega_{s}$ and $\Omega_{a}$ below the threshold of $0.25\gamma$, $\Omega_{s}= 0.21\gamma$ and $\Omega_{a}=0.033\gamma$. The atoms were initially in the state $\ket{\Psi(0)} =\ket{e_{1}}\ket{g_{2}}$. }
\label{fig4n}
\end{figure}

Figure~\ref{fig4n} shows the time evolution of the populations $P_{1}(t)=|C_{1}(t)|^{2}$ and $P_{2}(t) = |C_{2}(t)|^{2}$ for both $\tilde{\Omega}_{s}$ and $\tilde{\Omega}_{a}$ below threshold. At early times, the population $P_{1}(t)$ decreases whereas $P_{2}(t)$ increases until the populations become equal. At that time, the populations began to decay monotonically. The rate they decay is equal to $\gamma^{-}_{a}$,  the slowest decay rate of the antisymmetric state. Since the atoms are very strongly coupled to each other, $U(x_{21},z_{0})=0.95$, the decay rate $\gamma^{-}_{a}\approx 0.002\gamma$. Consequently, the effective decay time of the populations can be very long. The decay of the populations is irreversible so the evolution of the system is Markovian.

\subsection{$\Omega_{s}$ above threshold and $\Omega_{a}$ below threshold}

At large values of $U(x_{21},z_{0})$ it may happen that $\Omega_{a}< u/4$ and $\Omega_{s}> u/4$ even if the atoms are weakly coupled to the plasma field, i.e. $\Omega_{0} < u/4$. For example, when $U(x_{21},z_{0})\approx 1$, we have $\Omega_{a}\approx 0$ and $\Omega_{s}\approx 2\Omega_{0}$. Hence, $\Omega_{s}$ can be larger than $u/4$ even if $\Omega_{0} < u/4$ and at the same time $\Omega_{a}$ can be smaller than $u/4$. For $\Omega_{s}>u/4$ the Rabi frequency $\tilde{\Omega}_{s}$ is purely imaginary, and then the time evolution of the probability amplitude $C_{s}(t)$ takes the form
\begin{equation}
C_{s}(t) = C_{s}(0)e^{-\frac{1}{4}\left(\gamma +2i\delta\right)t}\left[\cos\bar{\Omega}_{s} t + \frac{(\gamma + 2i\delta)}{4\bar{\Omega}_{s}}\sin\bar{\Omega}_{s} t\right] , \label{a45n}
\end{equation}
where $\bar{\Omega}_{s} = \sqrt{\Omega_{s}^{2}-\frac{1}{16}(ue^{i\theta})^{2}}$. The temporal evolution of the $C_{s}(t)$ is sinusoidal whereas the temporal evolution of the amplitude $C_{a}(t)$, which is below threshold, is exponential and is given in Eq.~(\ref{a43}). In this case, the symmetric mode evolves in the non-Markovian regime whereas the antisymmetric mode evolves in the Markovian regime. It follows that in this case each atom evolves under the simultaneous influence of Markovian and non-Markovian mechanisms.
\begin{figure}[h]
\centering\includegraphics[width=9cm,keepaspectratio,clip]{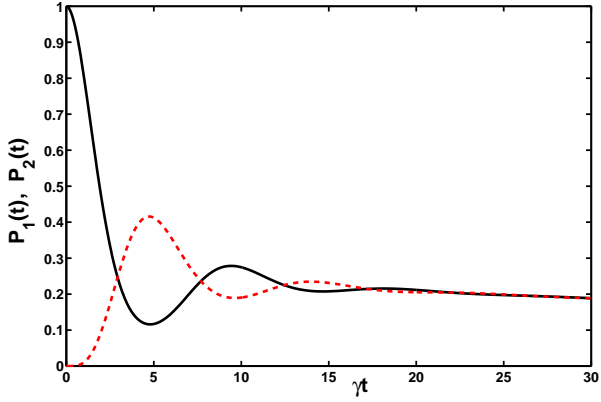}
\caption{(Color online) Time evolution of the populations $P_{1}(t)=|C_{1}(t)|^{2}$ (black solid line) and $P_{2}(t)=|C_{2}(t)|^{2}$ (red dashed line) for $\delta=0$, $\Omega_{0} = 0.5\gamma$, and $U(x_{21},z_{0})=0.99$ corresponding to $\Omega_{s}$ above and $\Omega_{a}$ below the threshold of $0.25\gamma$, $\Omega_{s}= 0.705\gamma$ and $\Omega_{a}=0.05\gamma$. The atoms were initially in the state $\ket{\Psi(0)} =\ket{e_{1}}\ket{g_{2}}$. }
\label{fig5n}
\end{figure}

Figure~\ref{fig5n} shows the evolution of the populations for $\Omega_{s}$ above and $\Omega_{a}$ below the threshold of $\frac{1}{4}\gamma$.
At early times, the oscillations of the populations with the Rabi frequency of the symmetric mode are clearly visible. At early times the populations oscillate with the Rabi frequency of the symmetric mode. In other words, the evolution of the populations is reversible but the reversibility occurs in a restricted time range $t < 1/\gamma^{-}_{a}$. Beyond $t\sim 1/\gamma^{-}_{a}$ the populations decay monotonically that the evolution is irreversible.
Thus, we can clearly distinguish between the non-Markovian and Markovian regimes of the evolutions. We see that the upper limit on time of the reversible evolution results from the presence of the interaction between the atoms. Clearly, it is a collective effect. Physically, it is a consequence of the fact that a large part of the population is trapped in the asymmetric state determined by the amplitude $\tilde{G}_{a}(t)$ thereby lowering the strength of the coupling of the atoms to the plasma field. It is easy to see, since for small distances between the atoms  $U(x_{21},z_{0})\approx 1$, we have $\Omega_{a}\approx 0$, which means that the antisymmetric states decouple from the plasma field. This example also shows that the atoms when behaving collectively can be weakly coupled  to the field even if individually they are strongly coupled to the field.

\subsection{Both $\Omega_{s}$ and $\Omega_{a}$ above threshold}

 Above the thresholds, $\tilde{\Omega}_{s}$ and $\tilde{\Omega}_{a}$ are purely imaginary. The time evolution of the probability amplitudes is then given by
\begin{align}
C_{s}(t) &= C_{s}(0)e^{-\frac{1}{4}\left(\gamma +2i\delta\right)t}\!\left[\cos\bar{\Omega}_{s} t + \frac{(\gamma + 2i\delta)}{4\bar{\Omega}_{s}}\sin\bar{\Omega}_{s} t\right] , \label{a45}\\
C_{a}(t) &= C_{a}(0)e^{-\frac{1}{4}\left(\gamma +2i\delta\right)t}\!\left[\cos\bar{\Omega}_{a} t + \frac{(\gamma + 2i\delta)}{4\bar{\Omega}_{a}}\sin\bar{\Omega}_{a} t\right] ,\label{a46}
\end{align}
where $\bar{\Omega}_{a} = \sqrt{\Omega_{a}^{2}-\frac{1}{16}(ue^{i\theta})^{2}}$. In this case, the time evolution of the probability amplitudes becomes sinusoidal. Such dynamics reflect the reversible property of the system that the evolution is non-Markovian.
\begin{figure}[h]
\centering\includegraphics[width=9cm,keepaspectratio,clip]{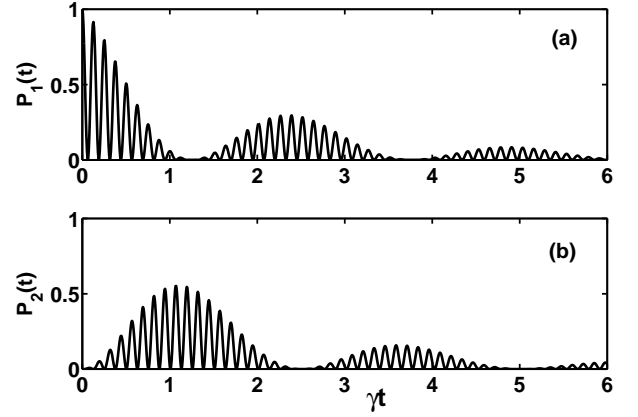}
\caption{Time evolution of the populations (a) $P_{1}(t)=|C_{1}(t)|^{2}$ and (b) $P_{2}(t)=|C_{2}(t)|^{2}$ for $\delta=0$, $\Omega_{0} =25\gamma$, and $U(x_{21},z_{0})=0.1$. The atoms were initially in the state $\ket{\Psi(0)} =\ket{e_{1}}\ket{g_{2}}$. }
\label{fig5}
\end{figure}

Figure~\ref{fig5} shows the time evolution of the populations $P_{1}(t)=|C_{1}(t)|^{2}$ and $P_{2}(t)=|C_{2}(t)|^{2}$ when the atoms are strongly coupled to the plasma field with $z=0.05\lambda_{s}$, $\delta =0$, but are weakly coupled to each other, $U(x_{21},z_{0})=0.1$. Note the presence of two characteristic time scales of the oscillations associated with the presence of two slightly different Rabi frequencies. At short times, $t\ll 1/\bar{\Omega}_{a}$, the initially excited atom $1$ periodically exchange the excitation with the plasma field at the Rabi frequency $\bar{\Omega}_{s}$. The population of the atom $2$ builds up with the oscillation of frequency $\bar{\Omega}_{s}$. The amplitudes of the populations are modulated with frequency $\bar{\Omega}_{a}$ causing collapses and revivals of the atomic populations.

\begin{figure}[h]
\centering\includegraphics[width=9cm,keepaspectratio,clip]{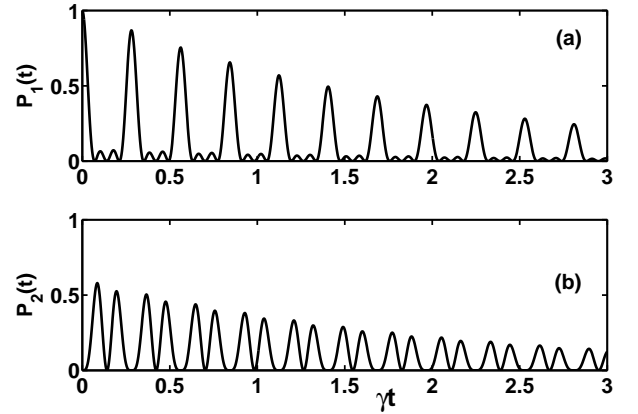}
\caption{Time evolution of the populations (a) $P_{1}(t)=|C_{1}(t)|^{2}$ and (b) $P_{2}(t)=|C_{2}(t)|^{2}$ for $\delta=0$, $\Omega_{0}=25\gamma$, and $U(x_{21},z_{0})=0.8$. The atoms were initially in the state $\ket{\Psi(0)} =\ket{e_{1}}\ket{g_{2}}$. }
\label{fig6}
\end{figure}

When the atoms are close to each other the collapses and revivals of the populations are absent. Instead, a periodic localization of the excitation is observed. This is illustrated in
figure~\ref{fig6} which shows the evolution of the populations for a small distance between the atoms at which $U(x_{21},z_{0})=0.8$. The manner the populations oscillate is different for $P_{1}(t)$ and $P_{2}(t)$. We see a periodic localization of the excitation that even at long times the memory effects are still evident. The explanation of this feature follows from the observation that at small distances between the atoms, the time scale of the oscillations of the antisymmetric state is very large, approaches infinity when  $U(x_{21},z_{0})\rightarrow 1$. Therefore, the system effectively evolves with a single time scale determined by the Rabi frequency of the symmetric state, $t\sim 1/\Omega_{s}$.

\section{Evolution of entanglement between the atoms}\label{sec5}

Given the time evolution of the probability amplitudes, we now proceed to evaluate the concurrence, a measure of entanglement between two qubits~\cite{sl10,zf9}.
Following the definition of the concurrence, we find that in terms of the probability amplitudes, $C_{1}(t)$ and $C_{2}(t)$, the concurrence is given by an expression
\begin{eqnarray}
C(t) = 2\left|C_{1}(t)C_{2}^{\ast}(t)\right| .
\end{eqnarray}
In terms of the amplitudes of the symmetric and antisymmetric combinations, the concurrence can be written as
\begin{eqnarray}
C(t)  = \left|\left[C_{s}(t) - C_{a}(t)\right]\!\left[C^{\ast}_{s}(t) + C^{\ast}_{a}(t)\right]\right| .\label{e47}
\end{eqnarray}
A positive value of the concurrence, $C(t)>0$, indicates entanglement between the atoms, and $C(t)=1$ corresponds to maximally entangled atoms. It is clear from Eq.~(\ref{e47}) that the atoms are entangled whenever $C_{s}(t)\neq C_{a}(t)$. Otherwise, the atoms are separable.
Thus, to examine the occurrence of entanglement between the atoms we must look at differences in the evolution of the amplitudes $C_{s}(t)$ and $C_{a}(t)$. If initially, $C_{s}(0)=C_{a}(0)$, then according to the solutions Eqs.~(\ref{a42}) and (\ref{a43}), the amplitudes will evolve differently only if $U(x_{21},z_{0})\neq 0$. It then follows that the coupling between the atoms through the plasma field is necessary to create entanglement between the atoms from an initial separable state.
\begin{figure}[h]
\centering\includegraphics[width=9cm,keepaspectratio,clip]{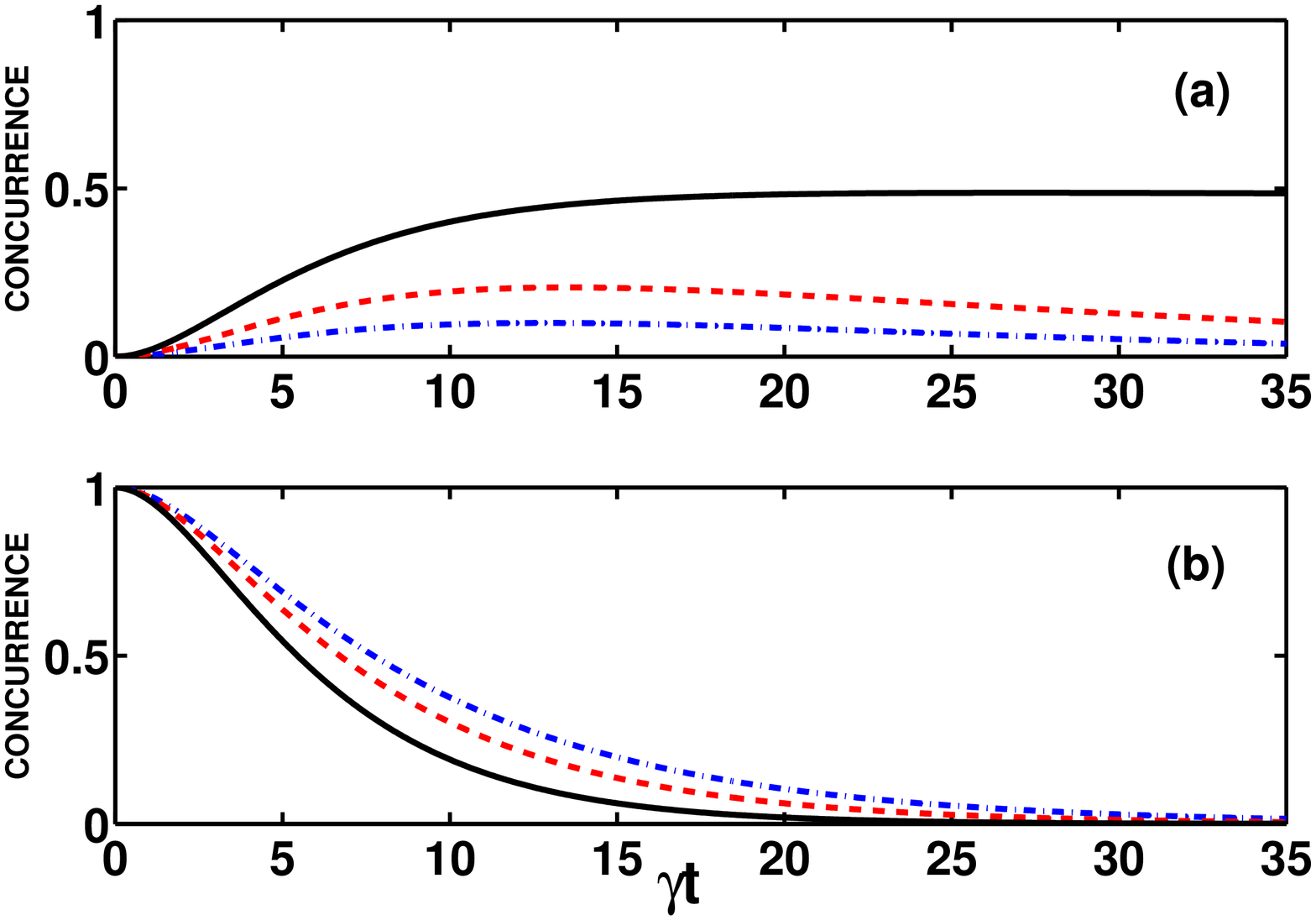}
\caption{(Color online) Concurrence versus time for the case of below threshold and two different initial states (a) $\ket{\Psi(0)} =\ket{e_{1}}\ket{g_{2}}$ and (b) $\ket{\Psi(0)} =\ket{s}$ with $\Omega_{0}=0.15\gamma$, $\delta =0$ and different $U(x_{21},z_{0})$: $U(x_{21},z_{0})=0.99$ (solid black line), $U(x_{21},z_{0})=0.5$ (dashed red line), $U(x_{21},z_{0})=0.25$ (dashed-dotted blue line).}
\label{fig9}
\end{figure}

\begin{figure}[h]
\centering\includegraphics[width=9cm,keepaspectratio,clip]{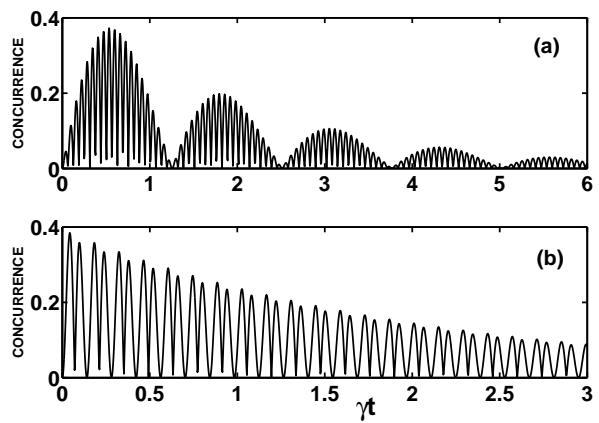}
\caption{Concurrence versus time for the case of above threshold  with $\Omega_{0}=25\gamma$, $\delta =0$, (a) $U(x_{21},z_{0})=0.1$ and (b) $U(x_{21},z_{0})=0.8$. The atoms were initially in the state $\ket{\Psi(0)} =\ket{e_{1}}\ket{g_{2}}$. }
\label{fig10}
\end{figure}

\begin{figure}[h]
\centering\includegraphics[width=9cm,keepaspectratio,clip]{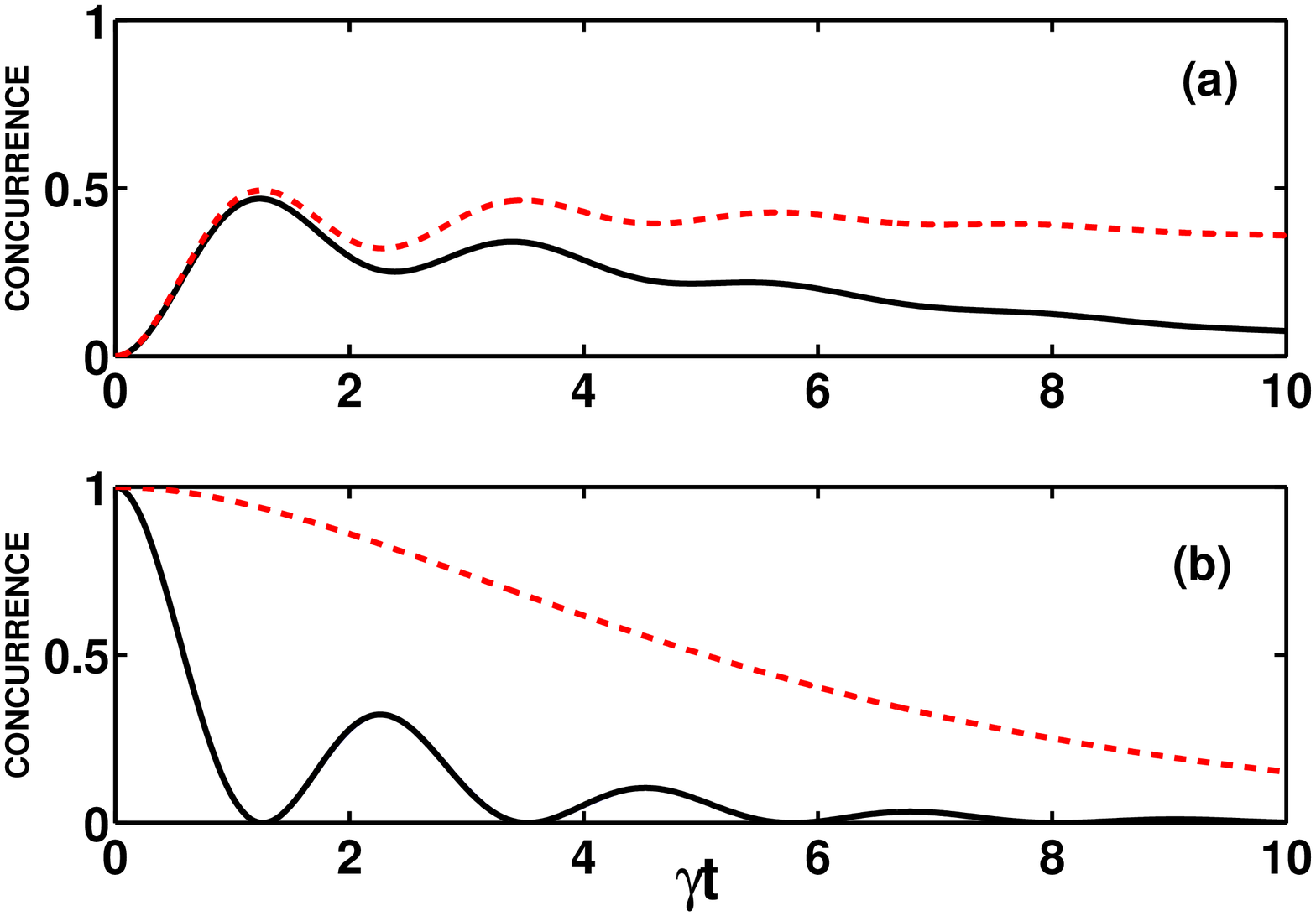}
\caption{(Color online) Concurrence versus time for $\Omega_{0}=\gamma$, $\delta =0$ and $U(x_{21},z_{0})\approx 1$ corresponding to the case of $\Omega_{s}$ above threshold but $\Omega_{a}$ below the threshold. Frame (a) shows the concurrence for $U(x_{21},z_{0})=0.95$ (solid black line) and $U(x_{21},z_{0})=0.99$ (dashed red line). The atoms were initially in a separable state $\ket{\Psi(0)} =\ket{e_{1}}\ket{g_{2}}$. Frame (b) shows the concurrence for $U(x_{21},z_{0})=0.95$ and two different initial states, the maximally entangled symmetric state $\ket{\Psi(0)} =\ket{s}$ (solid black line) and the maximally entangled antisymmetric state $\ket{\Psi(0)} =\ket{a}$ (dashed red line).}
\label{fig12}
\end{figure}

The features of the concurrence for the three regions of $\Omega_{s}$ and $\Omega_{a}$ are illustrated in Figs.~\ref{fig9} - \ref{fig12}.
Figure~\ref{fig9} shows the effect of increasing interaction strength between the atoms on the concurrence for a weak coupling of the atoms to the plasma field, both $\Omega_{s}$ and $\Omega_{a}$ below threshold, which corresponds to a Markovian evolution of the system. In figure~\ref{fig9}(a) the system starts from the separable state $\ket{e_{1}}\ket{g_{2}}$, whereas in figure~\ref{fig9}(b) the initial state of the system is the maximally entangled state $\ket s$. We see that even in the weak coupling regime, a large and long living entanglement can be created between the atoms. The entanglement created from the initial separable state increases with an increasing $U(x_{21},z_{0})$ and attains the maximal value of $C=0.5$ for $U(x_{21},z_{0})\approx 1$. The behavior of the concurrence is similar to that noted in the decay of two atoms into a common Markovian reservoir~\cite{zf9}.

When the system starts from a maximally entangled state, either $\ket s$ or $\ket a$, the initial entanglement always decays to zero with no entanglement present at long times, as illustrated in figure~\ref{fig9}(b). This is readily understood if it is recalled that the symmetric and antisymmetric states evolve independently in time. Thus, if the population of the antisymmetric state was initially zero it will remain zero for all times. In this case, the system of two atoms effectively behave as a single two-level system with the upper state $\ket s$ and the ground state $\ket g$. Then, the initial population of the state $\ket s$ decays exponentially to the ground state with the rate $2\Omega_{s}^{2}/\gamma$.

Turning now to the case of a strong coupling of the atoms to the plasma field at which $\Omega_{s}$ and $\Omega_{a}$ are above their thresholds, we show in figure~\ref{fig10} the evolution of the concurrence for weakly $(U(x_{21},z_{0})=0.1)$ and strongly $(U(x_{21},z_{0})=0.8)$ interacting atoms. The interaction creates a small difference between the frequencies of the oscillation of the symmetric and antisymmetric modes that $\bar{\Omega}_{s}\neq \bar{\Omega}_{a}$. The frequency difference induces beating oscillations of the populations of the atoms, as was seen in figure~\ref{fig5}, and one can see from figure~\ref{fig10} that these beating oscillations are rendered visible as beats in the concurrence.

Interesting features of the entanglement also appear when the symmetric mode evolves at Rabi frequency which is above the threshold, $\Omega_{s}>u/4$, and simultaneously the antisymmetric mode evolves at Rabi frequency which is below the threshold, $\Omega_{a}<u/4$. Under this circumstance, the probability amplitude of the symmetric mode is determined by Eq.~(\ref{a45}) whereas the amplitude of the antisymmetric mode is given by~Eq.~(\ref{a43}).

\begin{figure}[h]
\centering\includegraphics[width=9cm,keepaspectratio,clip]{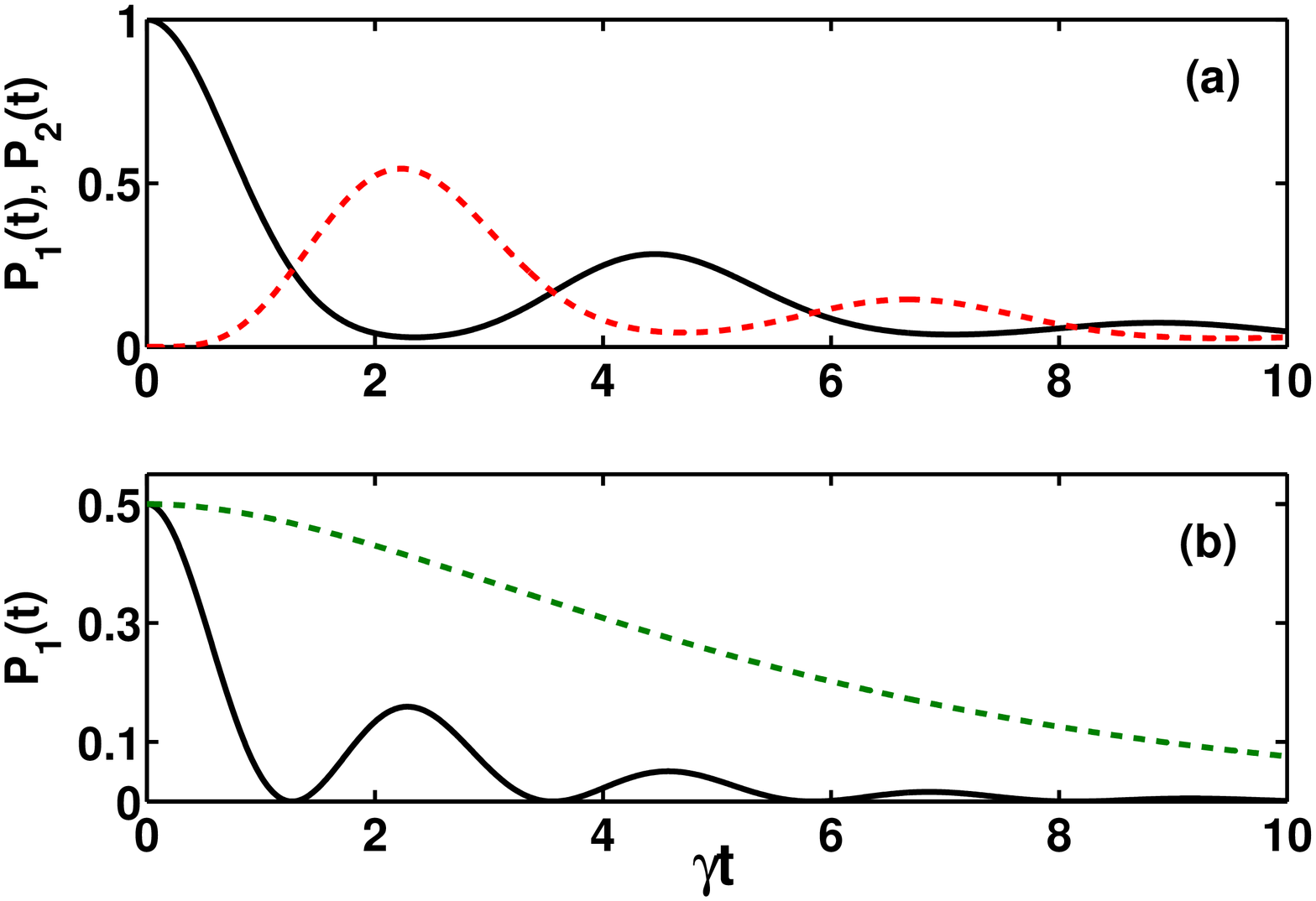}
\caption{(Color online) The time evolution of the populations of the atoms for the situation presented in figure~\ref{fig12}. Frame (a) shows the populations $P_{1}(t)$(solid black line) and $P_{2}(t)$ (dashed red line) for $\Omega_{0}=\gamma$, $\delta =0$ and $U(x_{21},z_{0})=0.95$. The atoms were initially in the state $\ket{\Psi(0)} =\ket{e_{1}}\ket{g_{2}}$. Frame (b) shows the time evolution of the population $P_{1}(t)$ for two different initial states, $\ket{\Psi(0)} =\ket{s}$ (solid black line) and $\ket{\Psi(0)} =\ket{a}$ (dashed green line). Not shown is $P_{2}(t)$ since in this case $P_{2}(t)= P_{1}(t)$.}
\label{fig13}
\end{figure}

Figure~\ref{fig12} shows the evolution of the concurrence for this special case. We see that the concurrence is zero only at the initial time $t=0$. As time progresses the concurrence develops to a nonzero value. The concurrence never becomes zero as time develops, and thus no periodic entanglement quenching occurs. This feature is associated with the fact that with the Rabi frequency $\Omega_{a}<u/4$, the population of the antisymmetric state does not evolve in time leading to a trapping of a part of the atomic populations in their energy states.
This is illustrated in figure~\ref{fig13}, which shows the time evolution of the population $P_{1}(t)$ and $P_{2}(t)$ for the same parameters as in figure~\ref{fig12}. We see from figure~\ref{fig13}(a) that the initial population is periodically transferred between the atoms. However, the transfer is not complete that the populations of the atoms never become zero during the evolution. A part of the population is trapped in the atoms and is not transferred between them. Figure~\ref{fig13}(b) shows the evolution of the population $P_{1}(t)$ for two initial maximally entangled states, $\ket s$ and $\ket a$. Since in this case $P_{2}(t)=P_{1}(t)$, we clearly see that the concurrence, if starts from maximally entangled state, it follows the evolution of the population of the atoms.

\section{Off-resonant coupling}\label{sec6}

To this end we have discussed the collective effects induced by the resonant interaction of the atoms with the plasma field. We have established the importance of the images of the atoms in the atomic dynamics. Moreover, we have demonstrated the equivalence of the system with that of four interacting atoms. We now turn to the off-resonant case of the atomic transition frequencies strongly detuned from the plasma frequency, $\delta \gg \gamma,\Omega_{s}$. Under such condition, the effective Rabi frequency $\tilde{\Omega}_{s}$ can be approximated~by
\begin{eqnarray}
\tilde{\Omega}_{s} \approx \frac{1}{4}\gamma\left(1-\frac{2\Omega_{s}^{2}}{\delta^{2}}\right) +\frac{1}{2}i\delta\left(1 +\frac{2\Omega_{s}^{2}}{\delta^{2}}\right)  .\label{f42}
\end{eqnarray}
Thus, if in Eq.~(\ref{a42}) the effective Rabi frequency is replaced by (\ref{f42}), we get, up to terms of order $\Omega_{s}^{2}/\delta^{2}$,
\begin{eqnarray}
C_{s}(t) \approx C_{s}(0)\!\left\{e^{-\left(\gamma -2i\delta\right)\frac{\Omega_{s}^{2}}{2\delta^{2}} t}+\frac{\Omega_{s}^{2}}{4\delta^{2}}e^{-\frac{1}{2}\left(\gamma +2i\delta\right) t}\right\} .\label{f44}
\end{eqnarray}
A similar expression with $s \rightarrow a$ gives $C_{a}(t)$. We see that $C_{s}(t)$ is composed of fast and slow oscillating terms varying in time with frequencies $\delta$ and $\Omega_{s}^{2}/\delta$, respectively. Of the two terms it is the one of the small magnitude $(\Omega_{s}^{2}/4\delta^{2})$ arising from the presence of the images. Thus, the evolution of $C_{s}(t)$ is well determined  without much contribution of the images. It is particularly well seen from Eq.~(\ref{c1}) that in the limit of $\delta \gg \Omega_{s}$, $\cos\phi\approx 1\, (\sin\phi\approx 0)$ so that the superposition amplitude $\tilde{D}_{s}(t)=0$ and $\tilde{D}_{a}(t)$ is reduced to the atomic amplitude $\tilde{C}_{s}(t)$.
In other words, two atoms significantly detuned from the plasma field are coupled each other by exchanging virtual photons through a short interaction time with the plasma field.
\begin{figure}[h]
\centering\includegraphics[width=9cm,keepaspectratio,clip]{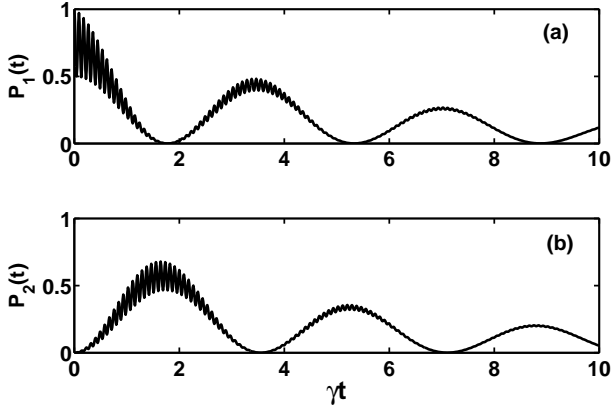}
\caption{Time evolution of the populations (a) $P_{1}(t)=|C_{1}(t)|^{2}$ and (b) $P_{2}(t)=|C_{2}(t)|^{2}$ for $\delta=50\gamma$, $\Omega_{0}=25\gamma$, and $U(x_{21},z_{0})=0.1$. The atoms were initially in the state $\ket{\Psi(0)} =\ket{e_{1}}\ket{g_{2}}$. }
\label{fig7}
\end{figure}

The above considerations are illustrated in figure~\ref{fig7}, which shows the time evolution of the atomic populations for $\delta =50\gamma$. We see that the atoms exchange the population with frequency $2\Omega_{0}^{2}/\delta$. The fast oscillations seen in the early time of the evolution occur at frequency $\delta$ and can be attributed to the involvement of the images in the dynamics of the system.
The presence of the fast oscillations only at very early times of the evolution is also consistent with energy-time uncertainty arguments. It is easy to understand. At short times the uncertainty of the energy of the atoms and the plasma field is very large, so one cannot distinguish between $\omega_{a}$ and $\omega_{s}$. This results in the presence of the fast oscillation of frequency $\delta$. As time progresses, the frequencies become more distinguishable resulting in the disappearance of the fast oscillations.

\begin{figure}[h]
\centering\includegraphics[width=9cm,keepaspectratio,clip]{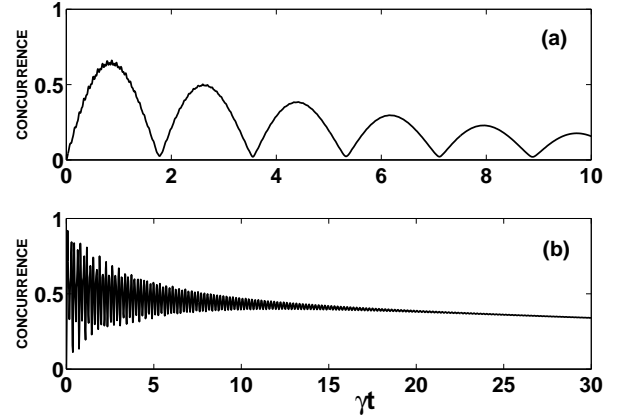}
\caption{Concurrence versus time for the case of above threshold  with $\Omega_{0}=25\gamma$, $\delta =50\gamma$ and different $U(x_{21},z_{0})$: (a) $U(x_{21},z_{0})=0.1$ and (b) $U(x_{21},z_{0})=0.95$. The atoms were initially in the state $\ket{\Psi(0)} =\ket{e_{1}}\ket{g_{2}}$. }
\label{fig11}
\end{figure}

It is interesting to contrast the entanglement created at $\delta\neq 0$ with that created in the resonant case of $\delta =0$.
We have seen in Sec.~\ref{sec5} that in the resonant case the maximal entanglement which can be created between the atoms from an initial separable state cannot exceed $C=0.5$. For off-resonant case $(\delta\neq 0)$, however, the entanglement can be significantly enhanced. This is shown in figure~\ref{fig11}, which illustrates the time evolution of the concurrence for a large detuning $\delta$. Clearly, at early times of the evolutions the concurrence is larger than~$1/2$, increases with an increasing $U(x_{21},z_{0})$ and becoming as large as $C=1$. This behavior can be explained in terms of the energy-time uncertainty relation. At short times a large uncertainty in the energy results in a large uncertainty in the localization of the excitation. We see that the increased possibility to distinguish between the frequencies of the atoms and the plasma field results in an enhanced entanglement between the atoms.

As a final remark, we would like to comment about a potential experimental system in which the collective dynamics of the atoms could be observed. It could be done in experiments similar
to that of Refs.~\cite{zz06,zz08,zz11}, where a strong coupling between an artificial atom embodied into a material structure composed of MN and EN meta-materials was observed. The strong coupling was observed as the Rabi oscillations of the temporal evolution of the electric field inside the artificial atom after being excited by a short pulse. The experimental setup could be modified by embodying two artificial atoms into the composed meta-material structure and observe the Rabi oscillation of the electric field of the atoms. The presence of the second atom could lead to the modulation of the Rabi oscillations of the population of the first atom, as seen in figures~\ref{fig4n}-\ref{fig6}, which would be the clear evidence of the collective behavior of the atoms.  

To clarify the role of the SPP in the collective behavior of the atoms, we now consider the emission properties near the interface of an atom, represented by its oscillating dipole ${\bf p}_{i}$, and calculate the electric field at position ${\bf r}$ emitted by the atom located at ${\bf r}_{i}$. The field is given by~\cite{M.S.Tomas}
\begin{eqnarray}
{\bf E}({\bf r}_{i}, {\bf r}, \omega_{a}) = \frac{\omega_{a}^{2}}{c^{2}}\stackrel{\leftrightarrow}{\bf G}({\bf r}_{i},{\bf r},\omega_{a})\cdot {\bf p}_{i} .
\end{eqnarray}
Suppose that the dipole moment ${\bf p}_{i}$ is polarized in the $x-z$ plane, ${\bf p}_{i} = p_{a}(\bar{{\bf x}}+\bar{{\bf z}})$. Then the coupling of atom $j$, located at an arbitrary position ${\bf r}$, to the field produced by the atom $i$ is maximal if the dipole moment ${\bf p}_{j}$ is parallel to ${\bf p}_{i}$. In this case, the magnitude of the field is
\begin{align}
&E({\bf r}_{i}, {\bf r}, \omega_{a}) = \frac{\omega_{a}^{2}}{c^{2}p_{a}}{\bf p}_{j}\cdot\stackrel{\leftrightarrow}{\bf G}({\bf r}_{i},{\bf r},\omega_{a})\cdot {\bf p}_{i} \nonumber\\
&= \frac{\omega_{a}^{2}p_{a}}{c^{2}}\left\{\Im[\stackrel{\leftrightarrow}{\bf G}\!({\bf r}_i,{\bf r},\omega)]_{xx}
+ \, \Im[\stackrel{\leftrightarrow}{\bf G}\!({\bf r}_i,{\bf r},\omega)]_{xz}\right. \nonumber\\
& +\left. \Im[\stackrel{\leftrightarrow}{\bf G}\!({\bf r}_i,{\bf r},\omega)]_{zx} + \Im[\stackrel{\leftrightarrow}{\bf G}\!({\bf r}_i,{\bf r},\omega)]_{zz}\right\}\nonumber\\
&= \frac{\hbar\varepsilon_{0}}{2p_{a}}\Omega^{2}_{0}U(x-x_{i},z_{i}) ,\label{eq45}
\end{align}
where $\Omega_{0}$ and $U(x-x_{i},z_{i})$ are given in Eqs.~(\ref{om0}) and (\ref{u}), respectively. Clearly, the function $U(x-x_{i},z_{i})$ determines the distribution of the field produced by atom $i$. It is seen from the expression (\ref{eq45}) that the field is distributed in the $x-z$ plane and the distribution depends not only on the distance $x-x_{i}$ along the interface but also on the distance $z_{i}$ of the radiating dipole from the interface. As illustrated in figure~\ref{fig4}, the variation of the function $U(x-x_{i},z_{i})$ with $x-x_{i}$ depends strongly on the distance $z_{i}\equiv z_{0}$ that $U(x-x_{i},z_{i})$, so that the field distribution,  decreases with an decreasing distance $z_{i}\equiv z_{0}$. Hence, $z_{0}$ should not be too small in order to achieve a strong coupling between distant atoms through their interaction with the SPP propagating along the interface.

\section{Conclusions}\label{sec7}

We have studied the dynamics of two two-level atoms located near to the interface of two meta-materials; one of negative permeability and the other of negative permittivity. We have derived analytical expressions for the probability amplitudes of the atomic states valid for an arbitrary initial state, arbitrary strengths of the coupling constants of the atoms to the plasma field, and arbitrary distances between the atoms. We have shown that the effect of the plasma field is to produce several interesting features, such as (1) two different time scales of the evolution of the atomic states, one corresponding to the evolution of the collective symmetric state and the other to the antisymmetric state. The existence of the two evolution time scales results in an entanglement between the atoms even in long times. (2) A threshold behavior of the coupling constants of the atoms to the plasma field which distinguishes between the non-Markovian and Markovian regimes of the evolutions. We have shown that the collective behavior of the atoms may lead to three different regimes of the evolution; fully Markovian, simultaneous Markovian and non-Markovian, and fully non-Markovian evolutions. The three regimes determines three different time scales of the evolution of the memory effects and entanglement. (3) In the case of the resonant coupling of the plasma field to the atoms,  the plasma field does not appear as a common reservoir to the atoms. We have adopted the image method and showed that in the resonant case the dynamics of the two atoms are completely equivalent to those of a four-atom system. (4) In the limit of a strong detuning of the plasma field from the atoms the dynamics resemble those of two atoms coupled to a common reservoir.

\section*{Acknowledgements}

This work is supported by the National Natural Science Foundation of China (Grant No. 61275123 and No.11474119) and the National Basic Research Program of China (Grant No. 2012CB921602).

\section*{Appendix A}\label{secA1}

In this Appendix we give details of the derivation of the integro-differential equations (\ref{e11u}) and (\ref{e12u}) for the probability amplitudes of the atomic states.

Equations of motion for the probability amplitudes are obtained from the Schr\"{o}dinger equation (\ref{es}), which with the interaction Hamiltonian (\ref{h3}) gives
\begin{align}
&\dot{C}_1(t) = -\frac{1}{\sqrt{\pi\varepsilon_{0}\hbar}}\int_{0}^{\infty} d\omega\, \frac{\omega}{c}e^{-i(\omega-\omega_{a})t} \nonumber\\
&\times \int d{\bf r}\left\{\frac{\omega}{c}\sqrt{\Im[\varepsilon({\bf r},\omega)]}{\bf p}_{1}\cdot\stackrel{\leftrightarrow}{\bf G}\!({\bf r}_1,{\bf r},\omega)\cdot{\bf C}_{e}({\bf r},\omega,t)\right. \nonumber\\
&+ \left. \sqrt{-\Im[\kappa({\bf r},\omega)]}{\bf p}_{1}\cdot[\stackrel{\leftrightarrow}{\bf G}\!({\bf r}_1,{\bf r},\omega)\times{\bf \nabla}]\cdot
{\bf C}_{m}({\bf r},\omega,t)\right\} ,\label{e8}\\
&\dot{C}_2(t) = -\frac{1}{\sqrt{\pi\varepsilon_{0}\hbar}}\int_{0}^{\infty} d\omega\, \frac{\omega}{c}e^{-i(\omega-\omega_{a})t} \nonumber\\
&\times \int d{\bf r} \left\{\frac{\omega}{c}\sqrt{\Im[\varepsilon({\bf r},\omega)]}
 {\bf p}_{2}\cdot\stackrel{\leftrightarrow}{\bf G}\!({\bf r}_2,{\bf r},\omega)\cdot{\bf C}_{e}({\bf r},\omega,t)\right. \nonumber\\
&+ \left. \sqrt{-\Im[\kappa({\bf r},\omega)]}{\bf p}_{2}\cdot[\stackrel{\leftrightarrow}{\bf G}({\bf r}_2,{\bf r},\omega)\times{\bf \nabla}]\cdot{\bf C}_{m}({\bf r},\omega,t)\right\} ,\label{e9}
\end{align}
and
\begin{align}
&\dot{{\bf C}}_{e}({\bf r},\omega,t) = \frac{1}{\sqrt{\pi\varepsilon_{0}\hbar}} \frac{\omega^{2}}{c^{2}}\sqrt{\Im[\varepsilon({\bf r},\omega)]}e^{i(\omega-\omega_{a})t}\nonumber\\
&\times\!\left[\stackrel{\leftrightarrow}{\bf G}^{\ast}\!\!\!({\bf r}_1,{\bf r},\omega)\cdot{\bf p}^{\ast}_{1}C_1(t)
+ \stackrel{\leftrightarrow}{\bf G}^{\ast}\!\!\!({\bf r}_2,{\bf r},\omega)\cdot{\bf p}^{\ast}_{2}C_2(t)\right] .\label{e10}\\
&\dot{{\bf C}}_{m}({\bf r},\omega,t) = \frac{1}{\sqrt{\pi\varepsilon_{0}\hbar}} \frac{\omega}{c}\sqrt{-\Im[\kappa({\bf r},\omega)]}e^{i(\omega-\omega_{a})t}\nonumber\\
&\times\!\left[{\bf \nabla}\times\!\stackrel{\leftrightarrow}{\bf G}^{\ast}\!\!\!({\bf r}_1,{\bf r},\omega)\!\cdot\!{\bf p}^{\ast}_{1}C_1(t)\!
+\! {\bf \nabla}\times\!\stackrel{\leftrightarrow}{\bf G}^{\ast}\!\!\!({\bf r}_2,{\bf r},\omega)\!\cdot\!{\bf p}^{\ast}_{2}C_2(t)\right] .\label{e10a}
\end{align}
We may eliminate the amplitudes for the field components by solving the equations for ${\bf C}_{e}({\bf r},\omega,t)$ and ${\bf C}_{m}({\bf r},\omega,t)$. Integrating Eqs.~(\ref{e10}) and (\ref{e10a}), and substituting the solutions into equations of motion for $C_{1}(t)$ and $C_{2}(t)$, we obtain
\begin{align}
\dot{C}_1(t) &= -\frac{1}{\pi\varepsilon_{0}\hbar c^{2}}\int^t_0dt'\int^\infty_0d\omega\, \omega^2 e^{-i(\omega-\omega_{a})(t-t')} \nonumber\\
&\times \left\{{\bf p}_{1}\cdot\Im[\stackrel{\leftrightarrow}{\bf G}({\bf r}_1,{\bf r}_1,\omega)]\cdot{\bf p}^{\ast}_{1}\, C_1(t')\right. \nonumber\\
&+\left. {\bf p}_{1}\cdot\Im[\stackrel{\leftrightarrow}{\bf G}({\bf r}_1,{\bf r}_{2},\omega)]\cdot{\bf p}^{\ast}_{2}\,C_2(t')\right\} ,\label{e11}\\
\dot{C}_2(t) &= -\frac{1}{\pi\varepsilon_{0}\hbar c^{2}}\int^t_0dt'\int^\infty_0d\omega\, \omega^2 e^{-i(\omega-\omega_{a})(t-t')}\nonumber\\
&\times \left\{{\bf p}_{2}\cdot\Im[\stackrel{\leftrightarrow}{\bf G}({\bf r}_{2},{\bf r}_{2},\omega)]\cdot{\bf p}^{\ast}_{2}\, C_2(t')\right. \nonumber\\
&+\left. {\bf p}_{2}\cdot\Im[\stackrel{\leftrightarrow}{\bf G}({\bf r}_2,{\bf r}_1,\omega)]\cdot{\bf p}^{\ast}_{1}\, C_1(t')\right\} ,\label{e12}
\end{align}
where we have used the following property of the Green's tensors~\cite{H.T.Dung}:
\begin{align}
&\int\!d{\bf r}\ \left\{\frac{\omega^{2}}{c^{2}}\Im[\varepsilon({\bf r},\omega)]\stackrel{\leftrightarrow}{\bf G}\!({\bf r}_{i},{\bf r},\omega)\stackrel{\leftrightarrow}{\bf G^{\ast}}\!({\bf r},{\bf r}_{j},\omega)-\right. \nonumber\\
&\left. \Im[\kappa({\bf r},\omega)][\stackrel{\leftrightarrow}{\bf G}\!({\bf r}_{i},{\bf r},\omega)\times\overleftarrow{\nabla}][\overrightarrow{\nabla}\times\stackrel{\leftrightarrow}{\bf G}^{\ast}\!({\bf r},{\bf r}_{j},\omega)]\right\} \nonumber\\
&= \Im[\stackrel{\leftrightarrow}{\bf G}\!({\bf r}_{i},{\bf r}_{j},\omega)] ,
\end{align}
in which $\Im[\stackrel{\leftrightarrow}{\bf G}({\bf r}_{i},{\bf r}_{j},\omega)]$ is the imaginary part of $\stackrel{\leftrightarrow}{\bf G}({\bf r}_{i},{\bf r}_{j},\omega)$.

If we introduce the notation
\begin{eqnarray}
K_{ij}(t,t^{\prime}) &=& -\frac{1}{\pi\varepsilon_{0}\hbar c^{2}}\int^\infty_0d\omega\, \omega^2 e^{-i(\omega-\omega_{a})(t-t')}\nonumber\\
&\times& {\bf p}_{i}\cdot\Im[\stackrel{\leftrightarrow}{\bf G}({\bf r}_i,{\bf r}_j,\omega)]\cdot{\bf p}^{\ast}_{j} ,\  i,j =1,2 ,\label{e12w}
\end{eqnarray}
we then easily find that Eqs.~(\ref{e11}) and (\ref{e12}) simplify to Eqs.~(\ref{e11u}) and (\ref{e12u}).

\section*{Appendix B}

In this Appendix, we evaluate the integral kernels of the integro-differential equations (\ref{e11u}) and (\ref{e12u}).
The kernels involve the imaginary part of the Green tensor $\stackrel{\leftrightarrow}{\bf G}\!({\bf r},{\bf r}_{i},\omega)$. The tensor when evaluated at an arbitrary space point ${\bf r}$, distance $R= |{\bf r}-{\bf r}_{i}|$ from the atom located at ${\bf r}_{i}$, can be written as~\cite{H.T.Dung}
\begin{eqnarray}
 \stackrel{\leftrightarrow}{\bf G}\!({\bf r},{\bf r}_{i},\omega) = \frac{c^2}{\omega^2\varepsilon_{1}}\left({\bf \nabla}{\bf \nabla}+k_{1}^{2}\stackrel{\leftrightarrow}{\mathbf{I}}\right)
\frac{e^{ik_{1}R}}{R} ,\label{w17}
\end{eqnarray}
where $k_{1}=\sqrt{\varepsilon_{1}\mu_{1}}\omega/c$ represents the wave number, $\varepsilon_{1}$ and $\mu_{1}$ are permittivity and permeability of the MN slab in which the atoms are located, and $\stackrel{\leftrightarrow}{\mathbf{I}}= \bar{\mathbf x}\bar{\bf x} + \bar{\bf y}\bar{\bf y}+ \bar{\bf z}\bar{\bf z}$ is the unit dyadic.
Using the Weyl's expansion~\cite{M.S.Tomas}
\begin{eqnarray}
\frac{e^{ik_{1}R}}{R} =\frac{i}{2\pi}\int d^{2}{\bf k}_{\parallel}\frac{e^{i\beta_{1}\mid z-z_{0}\mid}}{\beta_{1}}e^{i{\bf k}_{\parallel}\cdot({\bf r}-{\bf r}_{i})} ,
\end{eqnarray}
in which $\beta_{1}$ is the $z$ component of the propagation vector, we may express the Green tensor in terms of a two dimensional Fourier transform
\begin{align}
\stackrel{\leftrightarrow}{\bf G}\!({\bf r},{\bf r}_{i},\omega) &=\frac{1}{(2\pi)^2}\int d^{2}{\bf k}_{\parallel}\stackrel{\leftrightarrow}{\bf G}\!({\bf k}_{\parallel},\omega,z,z_{0})e^{i{\bf k}_{\parallel}\cdot({\bf r}- {\bf r}_{i})} ,\label{w19}
\end{align}
where $\stackrel{\leftrightarrow}{\bf G}({\bf k}_{\parallel},\omega,z,z_{0})$ is the Green tensor on a plane $({\bf k}_{\parallel},z)$ with constant $z$ coordinate, ${\bf k}_{\parallel} = k_{x}\bar{\bf x}+k_{y}\bar{\bf y}$ and ${\bf r}= r_{x}\bar{\bf x}+r_{y}\bar{\bf y}$ are components, respectively, of the wave vector and the position vector in a plane parallel to the interface, the $x-y$ plane.

The expression (\ref{w19}) allows us to evaluate the the Green tensor in terms of plane waves incident on and reflected from the boundaries between different materials, including the interface between the MN and EN materials and the boundaries between the materials and the exterior regions (vacuum) on either side. Since the atoms are located in the MN material, so that their radiative properties are modified by the field existing inside the material, we will evaluate the Green tensor only at points ${\bf r}$ inside the MN material. We follow the procedure of Toma\v{s}~\cite{M.S.Tomas} in evaluating the Green tensor.

The presence of the boundaries results in the field inside the MN material consisting of waves propagating in both the $+z$ and $-z$ directions. Therefore, $\stackrel{\leftrightarrow}{\bf G}({\bf k}_{\parallel},\omega,z,z_{0})$ can be written in terms of functions ${\bf U}_{q}^{\pm}({\bf k}_{\parallel},\omega,z)$ defined by imposing boundary conditions in the $z$ direction
\begin{align}
&\stackrel{\leftrightarrow}{\bf G}\!({\bf k}_{\parallel},\omega,z,z_{0}) = \frac{ i\mu_{1}}{2(2\pi)^{2}}\int d^{2}{\bf k_{\parallel}}\xi_{q}\frac{e^{i\beta_{1}d_{1}}}{\beta_{1}D_{q}}\nonumber\\
&\times \left[{\bf U}_{q}^{+}({\bf k}_{\parallel},\omega,z){\bf U}_{q}^{-}({\bf -k}_{\parallel},\omega,z_{0})\Theta(z-z_{0})\right. \nonumber\\
&+\left. {\bf U}_{q}^{-}({\bf k}_{\parallel},\omega,z){\bf U}_{q}^{+}({\bf -k}_{\parallel},\omega,z_{0})\Theta(z_{0}-z)\right]e^{i{\bf k_{\parallel}}({\bf \rho}-{\bf \rho_{0}})} ,
\end{align}
where the functions ${\bf U}_{q}^{\pm}({\bf k}_{\parallel},\omega,z)$ describe the electric field in MN slab, with unit strength incident from its upper side (by taking symbol '$-$') or lower side (by taking symbol '$+$'), that can be categorized into TM $(q=p)$ and TE $(q=s)$ types of even $(\xi_{p}=1)$ and odd $(\xi_{s} = -1)$ symmetries in the $\pm z$ directions, and $\Theta(z)$ is the unit step function. The forms of the functions ${\bf U}_{q}^{\pm}({\bf k}_{\parallel},\omega,z)$ for the field inside the MN material can be represented in terms of a sum of incident and reflected waves as
\begin{eqnarray}
{\bf U}_{q}^{+}({\bf k}_{\parallel},\omega,z) &=& {\bf e}^{+}_{q}({\bf k}_{\parallel})e^{i\beta_{1}(z-d_{1})} + r^{q}_{+}{\bf e}^{-}_{q}({\bf k}_{\parallel})e^{-i\beta_{1}(z-d_{1})} , \nonumber\\
{\bf U}_{q}^{-}({\bf k}_{\parallel},\omega,z) &=& {\bf e}^{-}_{q}({\bf k}_{\parallel})e^{-i\beta_{1}z} + r^{q}_{-}{\bf e}^{+}_{q}({\bf k}_{\parallel})e^{i\beta_{1}z}  ,
\end{eqnarray}
where ${\bf e}^{\pm}_{p}({\bf k}_{\parallel})=(\mp\beta_{1}{\bar{\bf k}}_{\parallel} +k_{\parallel}\bar{\bf z})/k_{1}$ and ${\bf e}^{\pm}_{s}({\bf k}_{\parallel})= \bar{\bf k}_{\parallel}\times\bar{\bf z}$ are orthonormal polarization vectors of the electric field of $p$ and $s$ polarized waves, respectively; $\bar{\bf k}_{\parallel}$ is the unit vector in the direction of ${\bf k}_{\parallel}$ $({\bf k}_{\parallel}=k_{\parallel}\bar{\bf k}_{\parallel})$, $\bar{\bf x}, \bar{\bf y}$ and $\bar{\bf z}$ are unit vectors in the Cartesian coordinates, $r^{q}_{\pm}$ are reflection coefficients of the waves propagating in the $\pm z$ directions, and
\begin{eqnarray}
D_{q} &=& 1-r_{-}^{q}r_{+}^{q}e^{2i\beta_{1}d_{1}}
\end{eqnarray}
results from summing the geometrical series due to the multiple reflections from the boundaries between different materials.

We now proceed to evaluate the components of the Green tensor in Cartesian coordinates, which are given by
\begin{equation}
G_{nm}({\bf r},{\bf r}_{i},\omega) = {\bf n}\cdot\left(\stackrel{\leftrightarrow}{\bf G}\!({\bf r},{\bf r}_{i},\omega)\right)\cdot{\bf m} ,
\end{equation}
where $n,m = x,y,z$, and ${\bf n, m} = \bar{\bf x}, \bar{\bf y}, \bar{\bf z}$.

Thus, if we apply the explicit forms of the polarization vectors, and use the polar representation for ${\bf k}_{\parallel}, {\bf k}_{\parallel} = k_{\parallel}(\cos\phi\bar{\bf x} +\sin\phi\bar{\bf y})$, the diagonal components of the Green tensor $\stackrel{\leftrightarrow}{\bf G}\!({\bf r},{\bf r}_{i},\omega)$ evaluated at a point ${\bf r}$ near the position ${\bf r}_{i}$ of the $i$th atom are then
\begin{align}
&G_{xx}({\bf r},{\bf r}_{i},\omega) = \frac{i\mu_{1}}{2(2\pi)^{2}}\!\!\int\!\!dk_\parallel\, k_\parallel \nonumber\\
&\times \int_{0}^{2\pi}\!\!d\phi \left[\frac{\beta_{1}\!\cos^{2}\!\phi}{k^2_1}R^{(p)}_{-}(z)
+\frac{\sin^{2}\!\phi}{\beta_{1}}R^{(s)}_{+}(z)\right]e^{i{\bf k}_{\parallel}\cdot ({\bf r}- {\bf r}_{i})} ,\nonumber\\
&G_{yy}({\bf r},{\bf r}_{i},\omega) = \frac{i\mu_{1}}{2(2\pi)^{2}}\int dk_\parallel\, k_\parallel \nonumber\\
&\times \int_{0}^{2\pi}\!\!d\phi\left[\frac{\beta_{1}\!\sin^{2}\!\phi}{k^2_1}R^{(p)}_{-}(z)
+\frac{\cos^{2}\!\phi}{\beta_{1}}R^{(s)}_{+}(z)\right]e^{i{\bf k}_{\parallel}\cdot ({\bf r}- {\bf r}_{i})} ,\nonumber\\
&G_{zz}({\bf r},{\bf r}_{i},\omega) = \frac{i\mu_{1}}{2(2\pi)^2}\!\int{\!dk_\parallel}\frac{k_\parallel^3 }{\beta_1k^2_1}R_{+}^{(p)}(z)\!\int_{0}^{2\pi}\!{d\phi}\, e^{i{\bf k}_{\parallel}\cdot ({\bf r}- {\bf r}_{i})} ,\label{e17}
\end{align}
and the off-diagonal components are
\begin{align}
&G_{xy}({\bf r},{\bf r}_{i},\omega) = G_{yx}({\bf r},{\bf r}_{i},\omega) =\frac{i\mu_{1}}{2(2\pi)^{2}}\!\!\int\!\!dk_\parallel\, k_\parallel \nonumber\\
&\times \int_{0}^{2\pi}\!\!d\phi \cos\phi\sin\phi\left[\frac{\beta_1}{k^2_1}R^{(p)}_{-}(z) -\frac{1}{\beta_{1}}R^{(s)}_{+}(z)\right]e^{i{\bf k}_{\parallel}\cdot ({\bf r}- {\bf r}_{i})} ,\nonumber\\
&G_{xz}({\bf r},{\bf r}_{i},\omega) = -G_{zx}({\bf r},{\bf r}_{i},\omega) =\frac{i\mu_{1}}{2(2\pi)^{2}}\!\!\int\!\!dk_\parallel\, k^{2}_\parallel \nonumber\\
&\times\!\!\int_{0}^{2\pi}\!d\phi \frac{\cos\phi}{k_{1}^{2}D_{p}}\!\left[r^{p}_{+}e^{-i\beta_1(z+z_0-2d_1)} -r^{p}_{-}e^{i\beta_1(z+z_0)}\right]\!e^{i{\bf k}_{\parallel}\cdot ({\bf r}- {\bf r}_{i})} ,\nonumber\\
&G_{yz}({\bf r},{\bf r}_{i},\omega) = -G_{zy}({\bf r},{\bf r}_{i},\omega) = \frac{i\mu_{1}}{2(2\pi)^{2}}\int dk_\parallel\, k^{2}_\parallel \nonumber\\
&\times\!\!\int\!d\phi \frac{\sin\phi}{k_{1}^{2}D_{p}}\bigg\{r^{p}_{+}e^{-i\beta_1(z+z_0-2d_1)} - r^{p}_{-}e^{i\beta_1(z+z_0)}\Bigg\}e^{i{\bf k}_{\parallel}\cdot ({\bf r}- {\bf r}_{i})} ,\label{e18}
\end{align}
where
\begin{align}
&R^{(q)}_{\pm}(z) = \frac{1}{D_{q}}\!\left[e^{i\beta_1(z-z_0)}
\pm r^{q}_{-}e^{i\beta_1(z+z_0)}\right. \nonumber\\
&\pm \left. r^{q}_{+}e^{-i\beta_1(z+z_0-2d_1)}\!+\! r^{q}_{+}r^{q}_{-}e^{-i\beta_1(z-z_{0}-2d_1)}\right] .\label{e18a}
\end{align}

\begin{figure}[h]
\centering\includegraphics[width=9.5cm,keepaspectratio,clip]{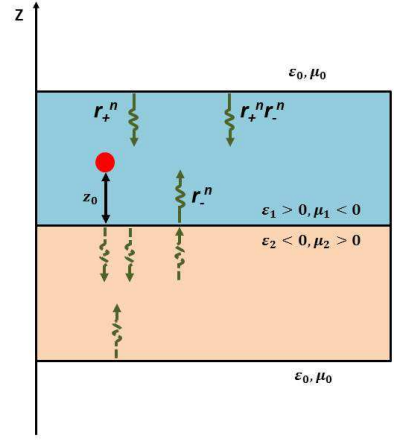}
\caption{(Color online) Illustration of reflected and transmitted waves propagating inside the MN and EN slabs. The red spots represents atoms located in the MN slab and emitting EM wave towards the interface of two materials.}
\label{fig2}
\end{figure}

The four terms appearing in Eq.~(\ref{e18a}) represent waves propagating inside the MN material. Figure~\ref{fig2} illustrates the source of those four terms in the expression (\ref{e17}).
The term $\exp[i\beta_{1}(z-z_{0})]$ represents a wave propagated a distance $z-z_{0}$ from the source atom located at $z_{0}$. This term is independent of the presence of the interface and the boundaries. Physically, it corresponds to the source field produced by an atom located at $z_{0}$.
The term $r_{-}\exp[i\beta_{1}(z+z_{0})]$ represents a wave propagated the distance $z+z_{0}$ after the reflection from the bottom interface of the MN material. The term $r_{+}\exp[-i\beta_{1}(z+z_{0}-2d_{1})]$ represents a wave propagated the distance $2d_{1}-z-z_{0}$ after the reflection from the upper interface of the MN material. The final term $r_{-}r_{+}\exp[-i\beta_1(z-z_0-2d_1)]$ represents a wave propagated the distance $2d_{1}+z_{0}-z$ after two reflections, one from the bottom and the other from the upper interfaces.

At this point it should be stressed that the expressions (\ref{e17}) and (\ref{e18}), although evaluated in the presence of the boundaries, they contain terms which are independent of the boundaries. The reason is in the fact that the field is not completely bounded into the area inside the materials. The slabs have finite sizes in the $z$ direction and in the derivation of Eqs.~(\ref{e17}) and (\ref{e18}) it has been assumed that there are nonzero transmission coefficients at the boundaries with the exterior vacuum regions.
Hence, we may consider the Green tensor as a sum of two terms, a source-field part $\stackrel{\leftrightarrow}{\bf G}_{S}({\bf r},{\bf r}_{i},\omega)$ and a scattered-field part $\stackrel{\leftrightarrow}{\bf G}_{B}({\bf r},{\bf r}_{i},\omega)$ as~\cite{H.T.Dung,M.S.Tomas}
\begin{eqnarray}
\stackrel{\leftrightarrow}{\bf G}\!\left({\bf r},{\bf r}_{i},\omega\right) =\, \stackrel{\leftrightarrow}{\bf G}_{S}\!\left({\bf r},{\bf r}_{i},\omega\right) + \stackrel{\leftrightarrow}{\bf G}_{B}\!\left({\bf r},{\bf r}_{i},\omega\right) .
\end{eqnarray}
In particular, $\stackrel{\leftrightarrow}{\bf G}_{S}({\bf r},{\bf r}_{i},\omega)$ is the Green tensor for the field which would exist in the material if there were no boundaries present, whereas $\stackrel{\leftrightarrow}{\bf G}_{B}({\bf r},{\bf r}_{i},\omega)$ is the Green tensor for the field scattered from the interface and boundaries. The source term has the same properties that would apply to the free (unbounded) field.

We may extract the source part $\stackrel{\leftrightarrow}{\bf G}_{S}\!\left({\bf r},{\bf r}_{i},\omega\right)$ from $\stackrel{\leftrightarrow}{\bf G}\!\left({\bf r},{\bf r}_{i},\omega\right)$ simply by putting $r^{q}_{\pm}=0$ in Eqs.~(\ref{e17}) and (\ref{e18}). This gives for the diagonal components
\begin{align}
G^{S}_{xx}({\bf r},{\bf r}_{i},\omega) &= \frac{i\mu_{1}}{2(2\pi)^{2}}\!\!\int\!\!dk_\parallel\, k_\parallel e^{i\beta_1(z-z_0)}\nonumber\\
&\times \int_{0}^{2\pi}\!\!d\phi \left[\frac{\beta_{1}\!\cos^{2}\!\phi}{k^2_1}
+\frac{\sin^{2}\!\phi}{\beta_{1}}\right]e^{i{\bf k}_{\parallel}\cdot ({\bf r}- {\bf r}_{i})} ,\nonumber\\
G^{S}_{yy}({\bf r},{\bf r}_{i},\omega) &= \frac{i\mu_{1}}{2(2\pi)^{2}}\int dk_\parallel\, k_\parallel e^{i\beta_1(z-z_0)}\nonumber\\
&\times \int_{0}^{2\pi}\!\!d\phi\left[\frac{\beta_{1}\!\sin^{2}\!\phi}{k^2_1}
+\frac{\cos^{2}\!\phi}{\beta_{1}}\right]e^{i{\bf k}_{\parallel}\cdot ({\bf r}- {\bf r}_{i})} ,\nonumber\\
G^{S}_{zz}({\bf r},{\bf r}_{i},\omega) &= \frac{i\mu_{1}}{2(2\pi)^2}\int{dk_\parallel}\frac{k_\parallel^3 }{\beta_1k^2_1} e^{i\beta_1(z-z_0)}\nonumber\\
&\times \int_{0}^{2\pi}{d\phi}\, e^{i{\bf k}_{\parallel}\cdot ({\bf r}- {\bf r}_{i})} ,\label{e17s}
\end{align}
and for the off-diagonal components
\begin{eqnarray}
G^{S}_{xy}({\bf r},{\bf r}_{i},\omega) &=& G^{S}_{yx}({\bf r},{\bf r}_{i},\omega) =\frac{i\mu_{1}}{2(2\pi)^{2}}\!\!\int\!\!dk_\parallel\, k_\parallel e^{i\beta_1(z-z_0)}\nonumber\\
&\times& \int_{0}^{2\pi}\!\!d\phi \cos\phi\sin\phi\left(\frac{\beta_1}{k^2_1} -\frac{1}{\beta_{1}}\right)e^{i{\bf k}_{\parallel}\cdot ({\bf r}- {\bf r}_{i})} ,\nonumber\\
G^{S}_{xz}({\bf r},{\bf r}_{i},\omega) &=& -G^{S}_{zx}({\bf r},{\bf r}_{i},\omega) = 0 ,\nonumber\\
G^{S}_{yz}({\bf r},{\bf r}_{i},\omega) &=& -G^{S}_{zy}({\bf r},{\bf r}_{i},\omega) = 0 .\label{e18f}
\end{eqnarray}

To evaluate the integrals over $\phi$, which appear in Eqs.~(\ref{e17}) and (\ref{e18}) and involving $\exp[i{\bf k}_{\parallel}\cdot ({\bf r}- {\bf r}_{i})]$, we assume, for simplicity, that ${\bf r}- {\bf r}_{i}$ has only $x$ component so that we can write the dot product in the form ${\bf k}_{\parallel}\cdot ({\bf r}- {\bf r}_{i}) = \alpha_{i}\cos\phi$, where $\alpha_{i} =k_{\parallel}|{\bf r}- {\bf r}_{i}|$. Hence, we arrive at the following expressions for the diagonal components
\begin{align}
&G_{xx}({\bf r},{\bf r}_{i},\omega) = \frac{i\mu_{1}}{4\pi}\!\!\int\!\!dk_\parallel\, k_\parallel \nonumber\\
&\times \left\{\frac{\beta_{1}}{k^2_1}\left[\frac{J_{1}(\alpha_{i})}{\alpha_{i}} - J_{2}(\alpha_{i})\right]R^{(p)}_{-}(z)
+\frac{J_{1}(\alpha_{i})}{\beta_{1}\alpha_{i}}R^{(s)}_{+}(z)\right\} ,\nonumber\\
&G_{yy}({\bf r},{\bf r}_{i},\omega) = \frac{i\mu_{1}}{4\pi}\int dk_\parallel\, k_\parallel \nonumber\\
&\times \left\{\frac{\beta_{1}J_{1}(\alpha_{i})}{k^2_1\alpha_{i}}R^{(p)}_{-}(z)
+\frac{1}{\beta_{1}}\left[\frac{J_{1}(\alpha_{i})}{\alpha_{i}} - J_{2}(\alpha_{i})\right]R^{(s)}_{+}(z)\right\} ,\nonumber\\
&G_{zz}({\bf r},{\bf r}_{i},\omega) = \frac{i\mu_{1}}{4\pi}\int{dk_\parallel}\frac{k_\parallel^3 }{\beta_1k^2_1}J_{0}(\alpha_{i})R_{+}^{(p)}(z) ,\label{e18p}
\end{align}
and for the off-diagonal components
\begin{align}
&G_{xy}({\bf r},{\bf r}_{i},\omega) = G_{yx}({\bf r},{\bf r}_{i},\omega) = 0 ,\nonumber\\
&G_{yz}({\bf r},{\bf r}_{i},\omega) = -G_{zy}({\bf r},{\bf r}_{i},\omega) = 0 ,\nonumber\\
&G_{xz}({\bf r},{\bf r}_{i},\omega) = -G_{zx}({\bf r},{\bf r}_{i},\omega) =\frac{-\mu_{1}}{4\pi}\!\!\int\!\!dk_\parallel\, k^{2}_{\parallel}J_{1}(\alpha_{i}) \nonumber\\
&\times\frac{1}{k_{1}^{2}D_{p}}\left[r^{p}_{+}e^{-i\beta_1(z+z_0-2d_1)} -r^{p}_{-}e^{i\beta_1(z+z_0)}\right] .\label{e18q}
\end{align}
For the source part, we get that only the diagonal elements are different from zero and are given by the following expressions
\begin{align}
&G^{S}_{xx}({\bf r},{\bf r}_{i},\omega) = \frac{i\mu_{1}}{4\pi}\!\!\int\!\!dk_\parallel\, k_\parallel \nonumber\\
&\times  \left\{\frac{\beta_{1}}{k^2_1}\left[\frac{J_{1}(\alpha_{i})}{\alpha_{i}} - J_{2}(\alpha_{i})\right]
+\frac{J_{1}(\alpha_{i})}{\beta_{1}\alpha_{i}}\right\}e^{i\beta_1(z-z_0)} ,\nonumber\\
&G^{S}_{yy}({\bf r},{\bf r}_{i},\omega) = \frac{i\mu_{1}}{4\pi}\!\!\int\!\!dk_\parallel\, k_\parallel \nonumber\\
&\times \left\{\frac{\beta_{1}J_{1}(\alpha_{i})}{k^2_1\alpha_{i}}
+\frac{1}{\beta_{1}}\!\left[\frac{J_{1}(\alpha_{i})}{\alpha_{i}} - J_{2}(\alpha_{i})\right]\right\}e^{i\beta_1(z-z_0)} ,\nonumber\\
&G^{S}_{zz}({\bf r},{\bf r}_{i},\omega) = \frac{i\mu_{1}}{4\pi}\int{dk_\parallel}\frac{k_\parallel^3 }{\beta_1k^2_1}J_{0}(\alpha_{i})e^{i\beta_1(z-z_0)} .\label{e18w}
\end{align}
We see that, in general, the expressions for the diagonal elements of the source part of the Green tensor have complex values. However, for a lossless negative index material with $\epsilon_{1} >0$ and $\mu_{1}<0$, the $z$ component of the propagation vector $\beta_{1}=\sqrt{\epsilon_{1}\mu_{1}}\omega/c$ is pure imaginary. It is easily verified that the resulting expressions, evaluated at $z=z_{0}$ are then real numbers, thereby leading to $\Im[\stackrel{\leftrightarrow}{\bf G}_{S}\!\left({\bf r}_{i},{\bf r}_{j},\omega\right)] =0$. In practice, the  material could posses some losses and then $\beta_{1}$ would not be a pure imaginary number. In our case, the losses in the materials are determined by the parameter $\gamma$. However, in typical materials the losses are small, $\gamma \approx 10^{-4}\omega_{s}$, and usually neglected.

Thus, for a material in which $\varepsilon$ and $\mu$ have opposite signs, the imaginary part of the Green tensor is solely determined by the imaginary part of $\stackrel{\leftrightarrow}{\bf G}_{B}\!\left({\bf r},{\bf r}_{i},\omega\right)$, which physically depicts the interaction between atoms and meta slabs. Therefore, we are left with the integral equations (\ref{e18p}) and (\ref{e18q}) to be evaluated. In order to evaluate the integrals we need explicit expressions of the reflection coefficients. They are determined from the Fresnel's law and the boundary conditions that we can easily obtain reflection coefficients of the surface between different materials
\begin{align}
r_{i\rightarrow j}^p = \frac{\beta_i\varepsilon_{j}\!-\!\beta_j\varepsilon_i}{\beta_i\varepsilon_{j}\!+\!\beta_j\varepsilon_i} ,\quad
r_{i\rightarrow j}^s = \frac{\beta_i\mu_{j}\!-\!\beta_j\mu_i}{\beta_i\mu_{j}\!+\!\beta_j\mu_i} ,\ i\neq j =\pm ,\label{e20}
\end{align}
where $i\rightarrow j$ indicates the direction of propagation of the wave, from the material $i$ to $j$. For a multiple-reflection case, the reflection coefficient is given by
\begin{eqnarray}
r_{i\rightarrow j\rightarrow k}^q=\frac{r_{i\rightarrow j}^{q}+r_{j\rightarrow k}^{q}e^{2i\beta_jd_j}}{1-r_{j\rightarrow i}^qr_{j\rightarrow k}^qe^{2i\beta_jd_j}}.\label{e21}
\end{eqnarray}

To evaluate the expression ${\bf p}_{i}\cdot\Im[\stackrel{\leftrightarrow}{\bf G}\!({\bf r}_i,{\bf r}_j,\omega)]\cdot{\bf p}^{\ast}_{j}$, we have to specify the orientation of the atomic dipole moments. If we assume that the atomic dipole moments are parallel to each other and are oriented in the $x-z$ plane
\begin{eqnarray}
{\bf p}_{1} = \frac{1}{\sqrt{2}}|{\bf p}_{1}|[1,0,1] ,\quad {\bf p}_{2} = \frac{1}{\sqrt{2}} |{\bf p}_{2}|[1,0,1] ,\label{w31}
\end{eqnarray}
we then find that the term ${\bf p}_{i}\cdot\Im[\stackrel{\leftrightarrow}{\bf G}\!({\bf r}_i,{\bf r}_j,\omega)]\cdot{\bf p}^{\ast}_{j}$ becomes
\begin{align}
&{\bf p}_{i}\cdot\Im[\stackrel{\leftrightarrow}{\bf G}\!({\bf r}_i,{\bf r}_j,\omega)]\cdot{\bf p}^{\ast}_{j} \nonumber\\
&= |{\bf p}_{i}||{\bf p}_{j}|\left(\bar{\bf x}+ \bar{\bf z}\right)\cdot\Im[\stackrel{\leftrightarrow}{\bf G}\!({\bf r}_i,{\bf r}_j,\omega)]\cdot(\bar{\bf x} + \bar{\bf z}) \nonumber\\
&= |{\bf p}_{i}||{\bf p}_{j}|\left\{\Im[\stackrel{\leftrightarrow}{\bf G}\!({\bf r}_i,{\bf r}_j,\omega)]_{xx}+\Im[\stackrel{\leftrightarrow}{\bf G}\!({\bf r}_i,{\bf r}_j,\omega)]_{xz}\right. \nonumber\\
&\left. +\, \Im[\stackrel{\leftrightarrow}{\bf G}\!({\bf r}_i,{\bf r}_j,\omega)]_{zx}+\Im[\stackrel{\leftrightarrow}{\bf G}\!({\bf r}_i,{\bf r}_j,\omega)]_{zz}\right\} .\label{w31}
\end{align}
Since $G_{xz}({\bf r}_{i},{\bf r}_{j},\omega) = -G_{zx}({\bf r}_{i},{\bf r}_{j},\omega)$, the term ${\bf p}_{i}\cdot\Im[\stackrel{\leftrightarrow}{\bf G}\!({\bf r}_{i},{\bf r}_j,\omega)]\cdot{\bf p}^{\ast}_{j}$ therefore becomes independent of the off-diagonal elements. Thus, with the choice of the orientation of the atomic dipole moments given by Eq.~(\ref{w31}) nonvanishing contributions to ${\bf p}_{i}\cdot\Im[\stackrel{\leftrightarrow}{\bf G}\!({\bf r}_{i},{\bf r}_j,\omega)]\cdot{\bf p}^{\ast}_{j}$ can come only from the diagonal $x$ and $z$ components of the Green tensor.

Although the choice of the dipole polarization in the $x-z$ plane affects the contribution of the diagonal elements of the Green tensor, it has no effect on the contribution of the off diagonal elements since independent of the atomic polarization all off-diagonal elements involving the $y$ component are zero.

The expressions for the components of the Green tensor, Eqs.~(\ref{e18p}), are exact and as such could be evaluated numerically for any values of the parameters involved. However, for many of the situations of interest the atoms will be located close to the interface, at a distance $z_{0}$ small compared to the radiation wavelength $\lambda_{a}=2\pi c/\omega_{a}$. Thus, we may limit ourselves then to considering the case, $z_{0} \ll \lambda_{a}$.
Furthermore, we may assume that  the thickness of the material slab is much larger than the localization length, $d_1\gg \lambda_{a}$. In this case, the factors $\exp[-i\beta_{1}(z\pm z_0 -2d_{1})]$ in the integrants of Eqs.~(\ref{e18p}) can be discarded leaving only the terms with factor $\exp[i\beta_{1}(z\pm z_0)]$ to contribute to these integrants. Moreover, in this limit the module of the term $\exp(2i\beta_{1}d_{1})$ in the expressions for $D_{s}$ and $D_{p}$ is much smaller than 1 and thus can also be neglected. We can then approximate $r_{i\rightarrow j\rightarrow k}^q$, as given in Eq.~(\ref{e21}), by $r_{i\rightarrow j}^q$ and when this result is inserted in Eq.~(\ref{e20}), we obtain for the reflection coefficients
\begin{eqnarray}
r_+^p &= r_{1\rightarrow0}^p=\frac{\beta_1\varepsilon_0-\beta_0\varepsilon_1}{\beta_1\varepsilon_0+\beta_0\varepsilon_1}
\approx \frac{\varepsilon_0-\varepsilon_1}{\varepsilon_0+\varepsilon_1} ,\label{e22}\\
r_-^p &= \frac{r_{1\rightarrow2}^p+r_{2\rightarrow0}^pe^{2i\beta_2d_2}}{1-r_{2\rightarrow0}^pr_{2\rightarrow1}^pe^{2i\beta_2d_2}}\approx \frac{\varepsilon_2-\varepsilon_1}{\varepsilon_1+\varepsilon_2} .\label{e23}
\end{eqnarray}
Before proceeding further we would like to point out that the reflection coefficient at the interface between the EN and MN materials, Eq.~(\ref{e23}), differs significantly from the reflection coefficient at the interface between an ordinary dielectric or a metal material~\cite{I.Avrutsky,E.N.Economou,A.Archambault}. According to Eq.~(\ref{e20}), the dispersion relation for TM-polarized mode propagating along the interface can be approximated by $\varepsilon_1\beta_2+\varepsilon_2\beta_1=0$. When $\varepsilon_1\mu_1=\varepsilon_2\mu_2$ and $\varepsilon_1=-\Re[\varepsilon_2]$, i.e. two slabs are perfectly paired, the dispersion relation reduces to $\varepsilon_1+\varepsilon_2=0$. In this case, the propagation of the plasma mode of frequency $\omega_s=\omega_{ep}/\sqrt{1+\varepsilon_1}$ is independent of the parallel component of the wave vector. This property is different from that of the plasma mode propagating at the interface of ordinary materials. In this case, the resonant plasma frequency depends on $k_\parallel$~\cite{I.Avrutsky}.

Hence, adopting the results of Eq.~(\ref{e1}), and assuming the frequency region of $\omega_{ep}\gg \omega \gg \omega_{eo}$, the reflection coefficients at the interface between the EN and MN slabs take the form
\begin{eqnarray}
r_-^p=1-\frac{\varepsilon_1\omega_s(\Delta\omega-\frac{1}{2}i\gamma)}{(\varepsilon_1+1)\left(\Delta\omega^{2}+\frac{1}{4}\gamma^2\right)} ,\label{e24}\\
r_-^s=-1+\frac{\mu_1\omega_s(\Delta\omega-\frac{1}{2}i\gamma)}{(\mu_1+1)(\Delta\omega^2+\frac{1}{4}\gamma^{2})} ,\label{e25}
\end{eqnarray}
where $\Delta\omega =\omega -\omega_s$ and, for simplicity, we have assumed that the dissipation parameters of the two slabs are equal, $\gamma_e=\gamma_m \equiv\gamma$.

If we now substitute Eqs.~(\ref{e24}) and (\ref{e25}) into Eqs.~(\ref{e18p}), we can perform the integration and arrive at the analytical expressions for the imaginary parts of the components of the Green tensor. Following the result (\ref{w31}), we evaluate only the diagonal $x$ and $z$ components of the Green tensor. Thus, when setting the parameter values $\varepsilon_{1}=\mu_2 =2$ and $\Re[\mu_1]=-\mu_2$, $\Re[\varepsilon_2]=-\varepsilon_1$, the explicit expressions for the imaginary parts of the diagonal $z$ component of the one- and two-point Green tensors are
\begin{eqnarray}
\Im[G_{zz}({\bf r}_{1},{\bf r}_{1},\omega)] = \frac{\gamma\omega_s}{12\pi k^{2}\left(\Delta\omega^2 +\frac{1}{4}\gamma^{2}\right)(2z_{0})^3} ,\label{w37}
\end{eqnarray}
and
\begin{align}
\Im[G_{zz}({\bf r}_{2},{\bf r}_{1},\omega)] &= \frac{\gamma\omega_s}{12\pi k^{2}\left(\Delta\omega^{2}+\frac{1}{4}\gamma^{2}\right)(2z_{0})^3} \nonumber\\
&\times F\left[\frac{3}{2},2,1;-\frac{x_{21}^2}{(2z_{0})^2}\right] .\label{w38}
\end{align}
where $k=\omega/c$, $x_{21}=x_{2}-x_{1}$ is the distance between the atoms, and $F(a,b,c;x)$ is the hypergeometrical function.

Similarly, for the diagonal $x$ component of the one- and two-point Green tensors, we find
\begin{align}
&\Im[G_{xx}({\bf r}_{1},{\bf r}_{1},\omega)] = \Im[G_{xx}({\bf r}_{2},{\bf r}_{2},\omega)] \nonumber\\
&=\frac{\gamma c^{2}\omega_{s}\!\left\{1-k_{s}^{2}\Re[\mu_1(\omega_s)](2z_{0})^2\right\}}{24\pi \omega^{2}\left(\Delta\omega^2 +\frac{1}{4}\gamma^{2}\right)(2z_{0})^3} ,\label{w39}\\
&\Im[G_{xx}({\bf r}_2,{\bf r}_1,\omega)] =  \Im[G_{xx}({\bf r}_{1},{\bf r}_{2},\omega)] \nonumber\\
&= \frac{\gamma c^{2}\omega_{s}}{24\pi \omega^{2}\left(\Delta\omega^2 +\frac{1}{4}\gamma^{2}\right)(2z_{0})^3} \nonumber\\
&\times \left\{\!F\!\left[\frac{3}{2},2,2;-\frac{x_{21}^2}{(2z_{0})^2}\right] -3\frac{x_{21}^2}{(2z_{0})^2}F\!\left[\frac{5}{2},3,3;-\frac{x_{21}^2}{(2z_{0})^2}\right]\right. \nonumber\\
&\left. -\Re[\mu_1(\omega_s)](2z_{0}k_{s})^{2} F\!\left[\frac{1}{2},1,2;-\frac{x_{21}^2}{(2z_{0})^2}\right]\right\} .\label{w40}
\end{align}
These expressions show that the imaginary parts of the Green tensor, when evaluated as a function of $\omega$, are of the form of a Lorentzian centered at the plasma frequency $\omega_s$ and possesses a bandwidth $\gamma/2$. Thus, for small $\gamma$ we expect that the largest contributions to the field come from $\omega\approx \omega_{s}$. Therefore, when substituting Eqs.~(\ref{w37})-(\ref{w40}) into Eq.~(\ref{e12w}), we can replace $\omega^{2}$ by $\omega_{s}^{2}\, (\approx \omega^{2}_{a})$ and extend the lower limit in the integration over $\omega$ to $-\infty$. Then after a simple algebra we obtain Eqs.~(\ref{ek1}) and~(\ref{ek2}).

\end{document}